\documentclass{emulateapj}

\usepackage{amsmath}
\usepackage{lscape}
\usepackage{multirow}

\newcommand{\about}{$\sim\!\!$~}

\newcommand{\be}{\begin{displaymath}}
\newcommand{\ee}{\end{displaymath}}

\def\lsim{\hbox{\rlap{\raise 0.425ex\hbox{$<$}}\lower 0.65ex\hbox{$\sim$}}}
\def\gsim{\hbox{\rlap{\raise 0.425ex\hbox{$>$}}\lower 0.65ex\hbox{$\sim$}}}

\def\arcdeg{\mbox{$^\circ$}}

\newcommand{\msun}{M$_\sun$}

\newcommand{\etal}{et al.\ }

\newcommand{\halpha}{H$\alpha$}

\newcommand{\hbeta}{H$\beta$}

\newcommand{\kms}{km~s$^{-1}$}
\newcommand{\ergps}{erg~s$^{-1}$}

\graphicspath{{/Users/jsilv/ptf11kx/plots/}}

\shorttitle{Supernovae Ia-CSM}
\shortauthors{Silverman, et al.}

\begin{document}

\title{Type Ia Supernovae Strongly Interacting with Their Circumstellar Medium}

\author{Jeffrey M. Silverman\altaffilmark{1,2,3}, Peter
  E. Nugent\altaffilmark{4}, Avishay Gal-Yam\altaffilmark{5},
  Mark Sullivan\altaffilmark{6}, D. Andrew Howell\altaffilmark{7,8},
  Alexei V. Filippenko\altaffilmark{2}, Iair Arcavi\altaffilmark{5},
  Sagi Ben-Ami\altaffilmark{5}, Joshua
  S. Bloom\altaffilmark{2}, S. Bradley Cenko\altaffilmark{2},
  Yi Cao\altaffilmark{9}, Ryan Chornock\altaffilmark{10}, Kelsey
  I. Clubb\altaffilmark{2}, Alison L. Coil\altaffilmark{11}, Ryan
  J. Foley\altaffilmark{9,12}, Melissa L. Graham\altaffilmark{7,8},
  Christopher V. Griffith\altaffilmark{13}, Assaf
  Horesh\altaffilmark{9}, Mansi M. Kasliwal\altaffilmark{14},
  Shrinivas R. Kulkarni\altaffilmark{9}, Douglas
  C. Leonard\altaffilmark{15}, Weidong Li\altaffilmark{2,16}, Thomas
  Matheson\altaffilmark{17}, Adam A. Miller\altaffilmark{2}, Maryam
  Modjaz\altaffilmark{18}, Eran O. Ofek\altaffilmark{5}, Yen-Chen
  Pan\altaffilmark{19}, Daniel A. Perley\altaffilmark{9}, Dovi
  Poznanski\altaffilmark{20}, Robert M. Quimby\altaffilmark{21}, Thea
  N. Steele\altaffilmark{2,22}, Assaf Sternberg\altaffilmark{23}, Dong
  Xu\altaffilmark{5}, and Ofer Yaron\altaffilmark{5}}

\altaffiltext{1}{Department of Astronomy, University of Texas, Austin, TX 78712-0259, USA.}
\altaffiltext{2}{Department of Astronomy, University of California, Berkeley, CA 94720-3411, USA.}
\altaffiltext{3}{email: jsilverman@astro.as.utexas.edu.}
\altaffiltext{4}{Lawrence Berkeley National Laboratory, Berkeley, CA 94720, USA.}
\altaffiltext{5}{Benoziyo Center for Astrophysics, The Weizmann Institute of Science, Rehovot 76100, Israel.}
\altaffiltext{6}{School of Physics and Astronomy, University of Southampton, Southampton, SO17 1BJ, UK.}
\altaffiltext{7}{Las Cumbres Observatory Global Telescope Network, Goleta, CA 93117, USA.}
\altaffiltext{8}{Department of Physics, University of California, Santa Barbara, CA 93106, USA.}
\altaffiltext{9}{Cahill Center for Astrophysics, California Institute of Technology, Pasadena, CA 91125, USA.}
\altaffiltext{10}{Harvard-Smithsonian Center for Astrophysics, Cambridge, MA 02138, USA.}
\altaffiltext{11}{Department of Physics, University of California, San Diego, La Jolla, CA 92093, USA.}
\altaffiltext{12}{Clay Fellow.}
\altaffiltext{13}{Department of Astronomy and Astrophysics, The Pennsylvania State University, University Park, PA 16802, USA.}
\altaffiltext{14}{Observatories of the Carnegie Institution of Science, Pasadena, CA 91101, USA.}
\altaffiltext{15}{Department of Astronomy, San Diego State University, San Diego, CA 92182-1221, USA.}
\altaffiltext{16}{Deceased 12 December 2011.}
\altaffiltext{17}{National Optical Astronomy Observatory, Tucson, AZ 85719-4933, USA.}
\altaffiltext{18}{New York University, Center for Cosmology and Particle Physics, Department of Physics, New York, NY 10003, USA.}
\altaffiltext{19}{Department of Physics (Astrophysics), University of Oxford, Keble Road, Oxford OX1 3RH, UK.}
\altaffiltext{20}{School of Physics and Astronomy, Tel Aviv University, Tel Aviv, Israel.}
\altaffiltext{21}{Kavli IPMU, The University of Tokyo, Kashiwanoha 5-1-5, Kashiwa, 277-8583, Japan.}
\altaffiltext{22}{Department of Computer Science, Kutztown University of Pennsylvania, Kutztown, Pennsylvania 19530, USA.}
\altaffiltext{23}{Max-Planck-Institut f\"{u}r Astrophysik, 85741 Garching, Germany.}

\begin{abstract}
Owing to their utility for measurements of cosmic acceleration,
Type Ia supernovae (SNe) are perhaps the best-studied class of SNe,
yet the progenitor systems of these explosions largely remain a
mystery. A rare subclass of SNe~Ia show evidence of strong interaction
with their circumstellar medium (CSM), and in particular, a
hydrogen-rich CSM; we refer to them as SNe~Ia-CSM. In the
first systematic search for such systems, we have identified 16 
SNe~Ia-CSM, and here we present new spectra of 13 of them. Six
SNe~Ia-CSM have been well-studied previously, three were
previously known but are analyzed in-depth for the first time here,
and seven are new discoveries from the Palomar Transient Factory.
The spectra of all SNe~Ia-CSM are dominated by \halpha\ emission (with 
widths of \about2000~\kms) and exhibit large H$\alpha$/H$\beta$ intensity 
ratios (perhaps due to collisional excitation of hydrogen via the SN ejecta
overtaking slower-moving CSM shells); moreover, they have an almost 
complete lack of \ion{He}{1} emission. They also show possible evidence
of dust formation through a decrease in the red wing of \halpha\
75--100~d past maximum brightness, and nearly all SNe~Ia-CSM
exhibit strong \ion{Na}{1}~D absorption from the host galaxy. The
absolute magnitudes (uncorrected for host-galaxy extinction) of SNe~Ia-CSM
are found to be $-21.3 \leq M_R \leq -19$~mag, and they also seem
to show ultraviolet emission at early times and strong infrared emission 
at late times (but no detected radio or X-ray emission). Finally, the host
galaxies of SNe~Ia-CSM are all late-type spirals similar to the 
Milky Way, or dwarf irregulars like the Large Magellanic Cloud, which
implies that these objects come from a relatively young stellar
population. This  work represents the most detailed analysis of the
SN~Ia-CSM class to date. 
\end{abstract}



\section{Introduction}\label{s:intro}

While Type Ia supernovae (SNe~Ia) have been used as precise distance
indicators for nearly two decades \citep{Phillips93}, the nature of
their progenitor systems and explosion mechanisms is still unclear
\citep[see][for further information]{Howell11}. While there is general
acceptance that they are the result of the thermonuclear explosion of
C/O white dwarfs (WDs), it is now likely that there are at least two
major channels that lead to a SN~Ia. The single-degenerate (SD)
channel occurs when the WD accretes matter from a nondegenerate
companion star \citep[e.g.,][]{Whelan73}, while the double-degenerate
(DD) channel is the result of the merger of two WDs
\citep[e.g.,][]{Iben84,Webbink84}.

Extremely nearby SNe~Ia which were discovered soon after explosion
have recently led to tight constraints on the size and luminosity of
the companion star, thus ruling out many plausible SD scenarios for
these objects
\citep[e.g.,][]{Nugent11,Ganeshalingam11,Brown12,Foley12,Bloom12,Silverman12:12cg}. In 
addition, the so-called super-Chandrasekhar mass SNe~Ia are thought to
contain $> 1.4$~\msun\ of SN ejecta and thus are likely formed from
the DD scenario
\citep[e.g.,][]{Howell06,Yamanaka09,Scalzo10,Silverman11,Taubenberger11}. However,
there are a few SNe~Ia that show strong evidence for the SD channel,
possibly with a red giant (RG) companion. Photoionization and
subsequent recombination of circumstellar medium (CSM) has been
observed in relatively normal SNe~Ia
\citep{Patat07,Blondin09,Simon09}, and CSM has been detected in the
spectra of at least 20\% of SNe~Ia in spiral galaxies
\citep{Sternberg11} and has been linked to SN~Ia explosion properties
\citep{Foley12:prog}. 

Even greater interaction with CSM has been seen in a small number of
SNe~Ia whose spectra contain strong, narrow hydrogen emission and whose
luminosities often exceed those of the more ``typical'' SNe~Ia that
follow the light-curve decline rate versus peak luminosity correlation 
\citep[i.e., the ``Phillips relation'';][]{Phillips93}. These ``hybrid''
objects resemble Type IIn SNe (SNe~IIn) and have been dubbed SNe~Ia/IIn,
Ian, IIa, and IIan. Under the standard SN classification scheme
\citep[e.g.,][]{Filippenko97}, any SN with hydrogen features in its
optical spectrum is considered a Type II SN and the subset of these
showing relatively narrow emission lines are referred to as SNe~IIn. 
Objects with relatively narrow hydrogen emission lines that are further 
linked (spectroscopically) to SNe~Ia could be denoted as ``Type IIna,''
though this moniker is somewhat cumbersome and obfuscating. Therefore,
in this work, we choose to label such events as ``SNe~Ia-CSM'' (or
sometimes as ``Ia-CSM objects'') to highlight the connection to the 
physically distinct Type Ia class.

SNe~Ia-CSM have spectra which appear to be ``diluted'' SN~Ia
spectra with the aforementioned narrow hydrogen emission lines
superposed (leading to the SN~IIn resemblance). Their light curves
are broad and quite long-lived, peaking at absolute magnitudes
brighter than about $-19$~mag. The two best-studied objects in this
class are SNe~2002ic and 2005gj
\citep{Hamuy03,Deng04,Kotak04,Wang04,Wood-Vasey04,Aldering06,Prieto07}. Until
recently there was some debate in the literature as to whether
these objects are truly SNe~Ia or are in fact some strange flavor of
core-collapse SN \citep[CCSN; e.g.,][]{Benetti06,Trundle08}.

This possible controversy seems to have been settled by the discovery
and analysis of PTF11kx (\citealt{Dilday12}; Silverman \etal
submitted). This object was discovered by the Palomar 
Transient Factory \citep[PTF;][]{Rau09,Law09} and shown to initially
resemble the somewhat overluminous Type Ia SN~1999aa
\citep{Li01:pec,Strolger02,Garavini04}. Optical spectra of PTF11kx soon
developed a strong \halpha\ feature with a P-Cygni profile, eventually
resembling spectra of SNe~2002ic and 2005gj, the previously mentioned
best-studied SNe~Ia-CSM. Using early-time data, \citet{Dilday12} 
find that PTF11kx was a {\it bona fide} SN~Ia with a symbiotic nova
progenitor (i.e., a SD scenario). Analyzing late-time data, (Silverman
\etal submitted) present further evidence that PTF11kx was
a SN~Ia that is strongly interacting with multiple thin CSM
shells. These findings can logically be extended to imply that 
{\it all} SNe~Ia-CSM are likely to be real SNe~Ia with
significant amounts of H-rich CSM, possibly caused in part by a
nondegenerate companion.

In \S\ref{s:iaCSM} we discuss our search for and identification (or
reidentification) of previously known SNe~Ia-CSM. Most of these have not 
been studied in much detail before, and we present new spectra of many of 
them. Similarly, in \S\ref{s:ptf_iaCSM}, we discuss new SNe~Ia-CSM
discovered by PTF. \S\ref{s:analysis} contains the analysis of our
spectra of all SNe~Ia-CSM presented herein, in addition to the
spectra of PTF11kx from \citet{Dilday12} and Silverman \etal
(submitted). Finally, we recap our conclusions in
\S\ref{s:conclusions} and summarize the major observational
characteristics shared by all SNe~Ia-CSM.


\section{Previously Known SNe~Ia-CSM}\label{s:iaCSM}

In order to find SNe~Ia-CSM that may have previously been
classified as SNe~IIn, we use the SuperNova IDentification code 
\citep[SNID;][]{Blondin07}. SNID classifies SN spectra by
cross-correlating an input spectrum with a large database of observed
SN spectra (known as ``templates''). To identify SNe~Ia-CSM, 
we created a special set of spectral templates that consisted of all
310 spectra of 163 objects classified as SNe~IIn from the Berkeley SN
Group's database \citep{Silverman12:BSNIPI}. In addition to these, we
also included templates of a handful of underluminous, overluminous,
and normal SNe~Ia at a variety of epochs, as well as 27 spectral
templates of the well-known Ia-CSM objects SNe~2002ic and 2005gj
(mentioned in \S\ref{s:intro}). 

We then ran all of our spectra of PTF11kx and the previously known
Ia-CSM objects SNe~1999E, 2002ic, and 2005gj (see
\S\S\ref{ss:02ic_05gj}, \ref{ss:97cy_99e}) through SNID using this
newly created template set, making sure to ignore any templates made
from the object currently under consideration. The best-matching
templates for each of these known SNe~Ia-CSM were visually
inspected, and SNe which were often found in the top 10--20
best-matching templates were flagged for further examination. This
yielded 18 possible SNe~Ia-CSM. Upon deeper analysis, it was found
that many of the spectra matched those of SNe~Ia-CSM and ``normal'' SNe~IIn 
equally well; also, spectra of some objects had very low signal-to-noise
ratios. While some of these may perhaps be true SNe~Ia-CSM, we cannot
confidently claim this for most of them. Thus, after this closer
inspection was conducted, four of the 18 possible SNe~Ia-CSM are
convincingly part of the SN~Ia-CSM class.

Interestingly, all four of our ``newly discovered'' members of the
SN~Ia-CSM class were noted to be similar to the prototypical Ia-CSM
objects SNe~2002ic and 2005gj, but only in unrefereed spectral 
classifications. From a literature search, no mention of these objects
as SNe~Ia-CSM is found anywhere else, except for SN~2008J which was
studied by \citet{Taddia12}, work that was made public during the
final stages of preparing this manuscript. Adding these four objects
to the prototypical SNe~2002ic and 2005gj and to the less-studied, but 
still fairly well known Ia-CSM objects SNe~1997cy and 1999E gives us a
sample size of eight SNe~Ia-CSM out of the hundreds of SNe~IIn discovered
in the last \about15~yr. Tables~\ref{t:iaCSM} and \ref{t:iaCSM_host} 
give basic information for all eight of these SNe~Ia-CSM and
their host galaxies, respectively, and our spectra of six of these SNe
are summarized in Table~\ref{t:iaCSM_spec}. Upon publication, all
spectra presented in this paper will be available in electronic format
on WISeREP \citep[the Weizmann Interactive Supernova data
REPository;][]{Yaron12}.\footnote{http://www.weizmann.ac.il/astrophysics/wiserep .}
These eight SNe~Ia-CSM are briefly discussed (in pairs) below. 

\begin{table*}
\scriptsize
\begin{center}
\caption{Non-PTF SNe~Ia-CSM}\label{t:iaCSM}
\begin{tabular}{lcccclc}
\hline
\hline
SN Name & Discovery & Approx. Date of & Discovery & Classification & \multicolumn{1}{c}{Peak Absolute} & Galactic \\
                   & Date & Maximum & Reference & Reference$^\textrm{a}$ & \multicolumn{1}{c}{Magnitude$^\textrm{b}$} & Reddening$^\textrm{c}$ \\
\hline
SN~1997cy & 1997~July~16 & $\cdots^\textrm{d}$ & IAUC 6706 & T00, G00 & $< -20.1$ ($V$) & 0.021 \\
SN~1999E & 1999~Jan.~15 & $\cdots^\textrm{d}$ & IAUC 7089 & IAUC 7090, R03& $< -19.5$ ($V$) & 0.088 \\
SN~2002ic & 2002~Nov.~13 & 2002~Nov.~27 & IAUC 8019 & IAUC 8028 & $< -20.3$ ($V$) & 0.059 \\
SN~2005gj & 2005~Sep.~26 & 2005~Oct.~14 & CBET 247 & CBET 302 & $\phantom{<} -20.2$ ($g$) & 0.121 \\
SN~2008J & 2008~Jan.~15 & 2008~Feb.~3 & CBET 1211 & CBET 1218 & $\phantom{<} -19.2$ ($r$) & 0.023 \\
SN~2008cg & 2008~May~5 & 2008~Apr.~29 & CBET 1366 & CBET 1420 & $\phantom{<} -19.4$ (unf) & 0.050 \\
SN~2011jb & 2011~Nov.~28 & $\cdots^\textrm{d}$ & CBET 2947 & CBET 2947 & $\phantom{<} -20.3$ ($R$) & 0.034 \\
CSS120327:110520--015205 & 2012~Mar.~27 & 2012~Mar.~8 & ATel 4081 & ATel 4081 & $\phantom{<} -20.5$ (unf) & 0.052 \\
\hline\hline
\multicolumn{7}{p{6.3in}}{$^\textrm{a}$ `T00' = \citet{Turatto00};
  `G00' = \citet{Germany00}; `R03' = \citet{Rigon03}.} \\
\multicolumn{7}{p{6.3in}}{$^\textrm{b}$The optical band of the peak
  absolute magnitude (and approximate date of maximum brightness) is
  given in parentheses; `unf' = unfiltered.} \\
\multicolumn{7}{p{6.3in}}{$^\textrm{c}$Galactic reddening toward each
  SN as derived from the dust maps of \citet{Schlegel98}; includes
  the corrections of \citet{Peek10}.} \\
\multicolumn{7}{p{6.3in}}{$^\textrm{d}$SN was discovered after maximum
  brightness.} \\
\hline\hline
\end{tabular}
\end{center}
\end{table*}
\normalsize

\begin{table*}
\begin{center}
\caption{Non-PTF Ia-CSM Host Galaxies}\label{t:iaCSM_host}
\begin{tabular}{lccc}
\hline
\hline
SN Name &Name & Type & Redshift $z$ \\
\hline
SN~1997cy & Sersic 040/06:[GGP90] 342 & Dwarf Irregular & 0.0642 \\
SN~1999E & GSC 6116 00964 & Late-Type Spiral & 0.0258 \\
SN~2002ic & NEAT J013002.81+215306.9 & Late-Type Spiral (Sbc) & 0.0660 \\
SN~2005gj & SDSS J030111.99$-$003313.9 & Dwarf Irregular & 0.0616 \\
SN~2008J & MCG -02-07-033 & Late-Type Spiral (SBbc) & 0.0159 \\
SN~2008cg & FGC 1965 & Late-Type Spiral (Scd) & 0.0362 \\
SN~2011jb & SDSS J113704.81+152813.9 & Dwarf Irregular & 0.0826 \\
CSS120327:110520--015205 & SDSS J110520.10--015204.9  & Dwarf Irregular & 0.0908 \\
\hline\hline
\end{tabular}
\end{center}
\end{table*}

\begin{table*}
\scriptsize
\begin{center}
\caption{Spectra of Non-PTF SNe~Ia-CSM}\label{t:iaCSM_spec}
\begin{tabular}{lcccrr}
\hline
\hline
UT Date &  Age (d)$^\textrm{a}$  &  Instrument$^\textrm{b}$ & Range (\AA)  & Res. (\AA)$^\textrm{c}$ & Exp. (s) \\
\hline
\multicolumn{6}{c}{SN~1999E$^\textrm{d}$} \\
\hline
1999~Jan.~19.7 & $\phantom{00}$4 & LRIS & 5120--8850$\phantom{1}$ & 7 & 200 \\
1999~Jan.~20.7 & $\phantom{00}$5 & LRIS & 3760--6240$\phantom{1}$ & 4.5 & 200 \\
1999~Jan.~21.7 & $\phantom{00}$6 & LRIS & 6450--10200 & 7 & 300 \\
1999~Feb.~12.5 & $\phantom{0}$27 & Kast & 3382--10500 & 6/11 & 1800 \\
1999~Feb.~23.5 & $\phantom{0}$38 & Kast & 3366--10550 & 6/11 & 1200 \\
1999~Mar.~12.5 & $\phantom{0}$55 & Kast & 3442--10466 & 6/11 & 1800 \\
\hline
\multicolumn{6}{c}{SN~2005gj$^\textrm{e}$} \\
\hline
2005~Dec.~2.4  & $\phantom{0}$46 & DEIMOS & 3897--9070$\phantom{1}$ & 3 & 300 \\
2005~Dec.~4.4  & $\phantom{0}$48 & LRIS   & 3280--9320$\phantom{1}$ & 4.5/7 & 300 \\
2006~Jan.~1.4  & $\phantom{0}$74 & DEIMOS & 3918--9061$\phantom{1}$ & 3 & 600 \\
2006~Dec.~23.4 & 409 & DEIMOS & 4496--9574$\phantom{1}$ & 3 & 1500 \\
2007~Feb.~14.3 & 459 & LRIS   & 3206--9238$\phantom{1}$ & 6.4/7 & 1200 \\
\hline
\multicolumn{6}{c}{SN~2008J} \\
\hline
2008~Feb.~16.2 & $\phantom{0}$12 & Kast & 4440--10500 & 4.9/11.9 & 900 \\
2008~Aug.~1.5  & 177 & Kast & 4292--10800 & 6.0/11.0 & 900 \\
2008~Aug.~26.5 & 201 & Kast & 4232--10800 & 6.7/12.3 & 900 \\
2008~Sep.~7.5  & 213 & Kast & 4050--10740 & 7.4/12.3 & 1200 \\
2008~Sep.~22.5 & 228 & Kast & 3488--10344 & 4.6/11.0 & 1200 \\
2008~Oct.~7.5  & 243 & Kast & 3554--10780 & 4.4/11.0 & 1200 \\
2008~Oct.~22.4 & 258 & Kast & 3476--6400$\phantom{1}$ & 9.0 & 2400 \\
2008~Nov.~20.3 & 286 & Kast & 3480--10000 & 5.0/11.0 & 1800 \\
2008~Nov.~23.4 & 289 & Kast & 3420--8000$\phantom{1}$ & 4.8/5.0 & 1200 \\
\hline
\multicolumn{6}{c}{SN~2008cg} \\
\hline
2008~May~8.4 & $\phantom{00}$9 & Kast & 3300--10800 & 5.4/11.6 & 1800 \\
2008~May~15.4 & $\phantom{0}$15 & Kast & 3300--10800 & 5.5/11.7 & 1800 \\
2008~June~29.3 & $\phantom{0}$59 & Kast & 3300--10790 & 5.9/10.8 & 1500 \\
2008~July~7.3 & $\phantom{0}$67 & Kast & 3306--10800 & 6.0/11.3 & 1500 \\
2008~Aug.~27.3 & 116 & LRIS & 3268--9240$\phantom{1}$ & 4.5/7.0 & 454 \\
\hline
\multicolumn{6}{c}{SN~2011jb} \\
\hline
2011~Dec.~24.4 & $\phantom{0}$24 & Kast & 3518--10138 & 4.7/9.8 & 2400 \\
2012~June~16.4 & 186 & LRIS & 3350--10152 & 3.6/6.1 & 600 \\
2012~July~16.3 & 213 & LRIS & 3727--9939$\phantom{1}$ & 3.7/6.4 & 600 \\
\hline
\multicolumn{6}{c}{CSS120327:110520--015205} \\
\hline
2012~May~17.2 & $\phantom{0}$64 & LRIS & 3343--10100 & 3.7/6.2 & 300 \\
2012~June~16.3 & $\phantom{0}$92 & LRIS & 3346--10069 & 4.1/6.1 & 1200 \\
\hline\hline
\multicolumn{6}{p{4.2in}}{$^\textrm{a}$Rest-frame days relative to
  maximum brightness, except for SNe~1999E and 2011jb where the epoch
  is relative to the UT date of discovery (1999~Jan.~15 and 2011~Nov.~28,
  respectively). See Table~\ref{t:iaCSM} for the dates of maximum
  brightness for the rest of the objects.} \\
\multicolumn{6}{p{4.2in}}{$^\textrm{b}$LRIS = Low Resolution Imaging
  Spectrometer on the Keck 10~m telescope; Kast = Kast double
  spectrograph on the Shane 3~m telescope at Lick Observatory; DEIMOS
  = DEep Imaging Multi-Object Spectrograph on the Keck 10~m
  telescope.} \\
\multicolumn{6}{p{4.2in}}{$^\textrm{c}$Approximate full width at
  half-maximum intensity (FWHM) resolution. If
  two numbers are listed, 
  they represent the blue-side and red-side resolutions, respectively.} \\ 
\multicolumn{6}{p{4.2in}}{$^\textrm{d}$These spectra of SN~1999E have
been previously published by \citet{Filippenko00}.} \\
\multicolumn{6}{p{4.2in}}{$^\textrm{e}$These spectra of SN~2005gj have
been previously published by \citet{Silverman12:BSNIPI}.} \\
\hline\hline
\end{tabular}
\end{center}
\end{table*}
\normalsize

\subsection{SN~2002ic and SN~2005gj}\label{ss:02ic_05gj}

\begin{figure*}
\centering
\includegraphics[width=6.8in]{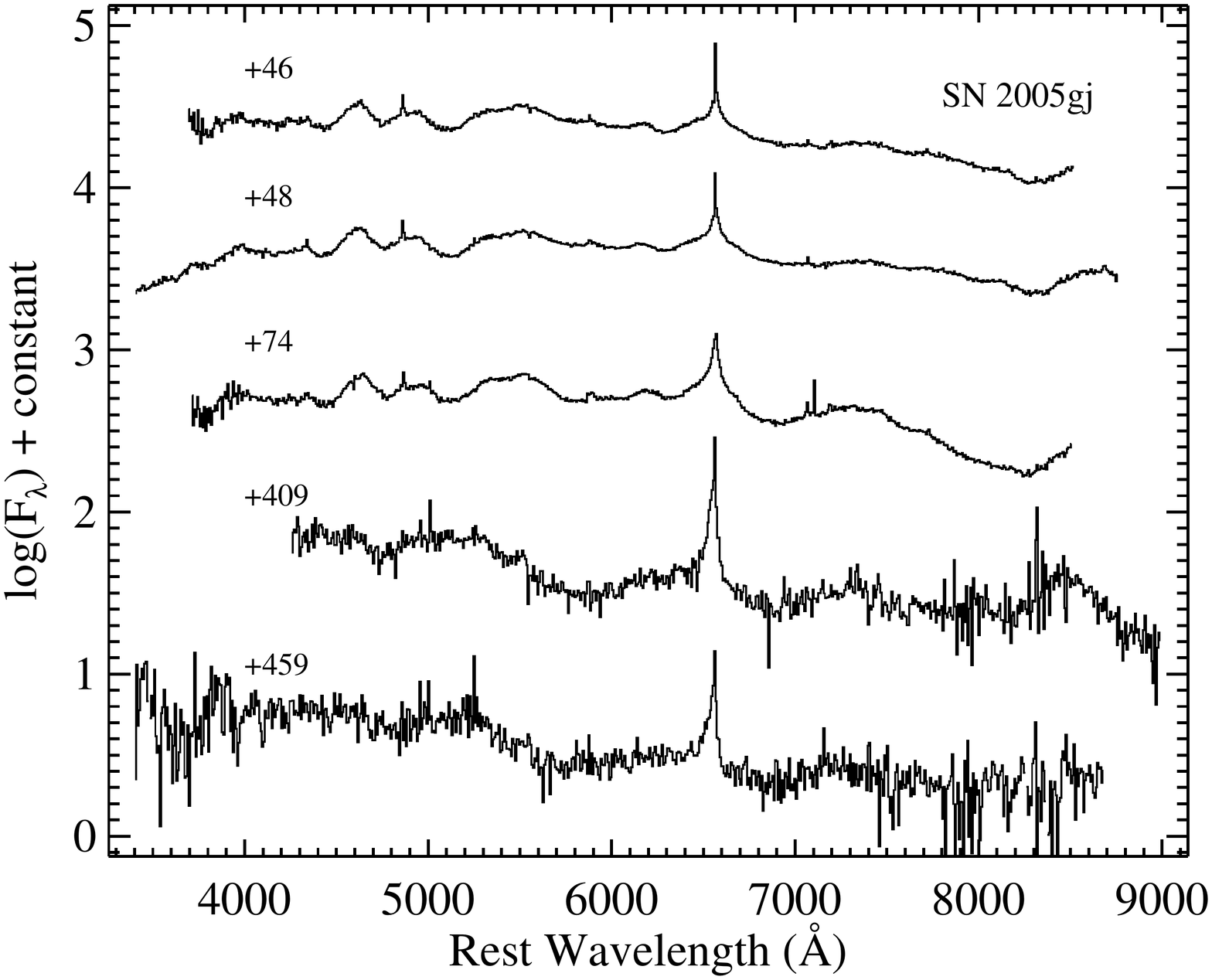}
\caption{Spectra of SN~2005gj, originally published by
  \citet{Silverman12:BSNIPI}, labeled with age relative to maximum
  brightness. The data have had their host-galaxy recession velocity 
  removed and have been corrected for Galactic
  reddening.}\label{f:sn2005gj}
\end{figure*}

As already mentioned, the two prototypical and
best-studied SNe~Ia-CSM are SN~2002ic
\citep{Hamuy03,Deng04,Kotak04,Wang04,Wood-Vasey04} and SN~2005gj
\citep{Aldering06,Prieto07}. The photometric and spectral evolution of
these two objects has been well-studied in the works cited above. Both
were found to be more luminous than typical SNe~Ia and most SNe~IIn,
have spectra consisting of relatively narrow \halpha\ emission (with P-Cygni
profiles) on top of a ``diluted'' SN~Ia spectrum, and broad,
slowly evolving light curves. To the already impressive data on
SN~2005gj in the literature, we add five spectra (three at early times and 
two at very late times) that were originally published by
\citet{Silverman12:BSNIPI} and are displayed in
Figure~\ref{f:sn2005gj}.

\subsection{SN~1997cy and SN~1999E}\label{ss:97cy_99e}

\begin{figure*}
\centering
\includegraphics[width=6.8in]{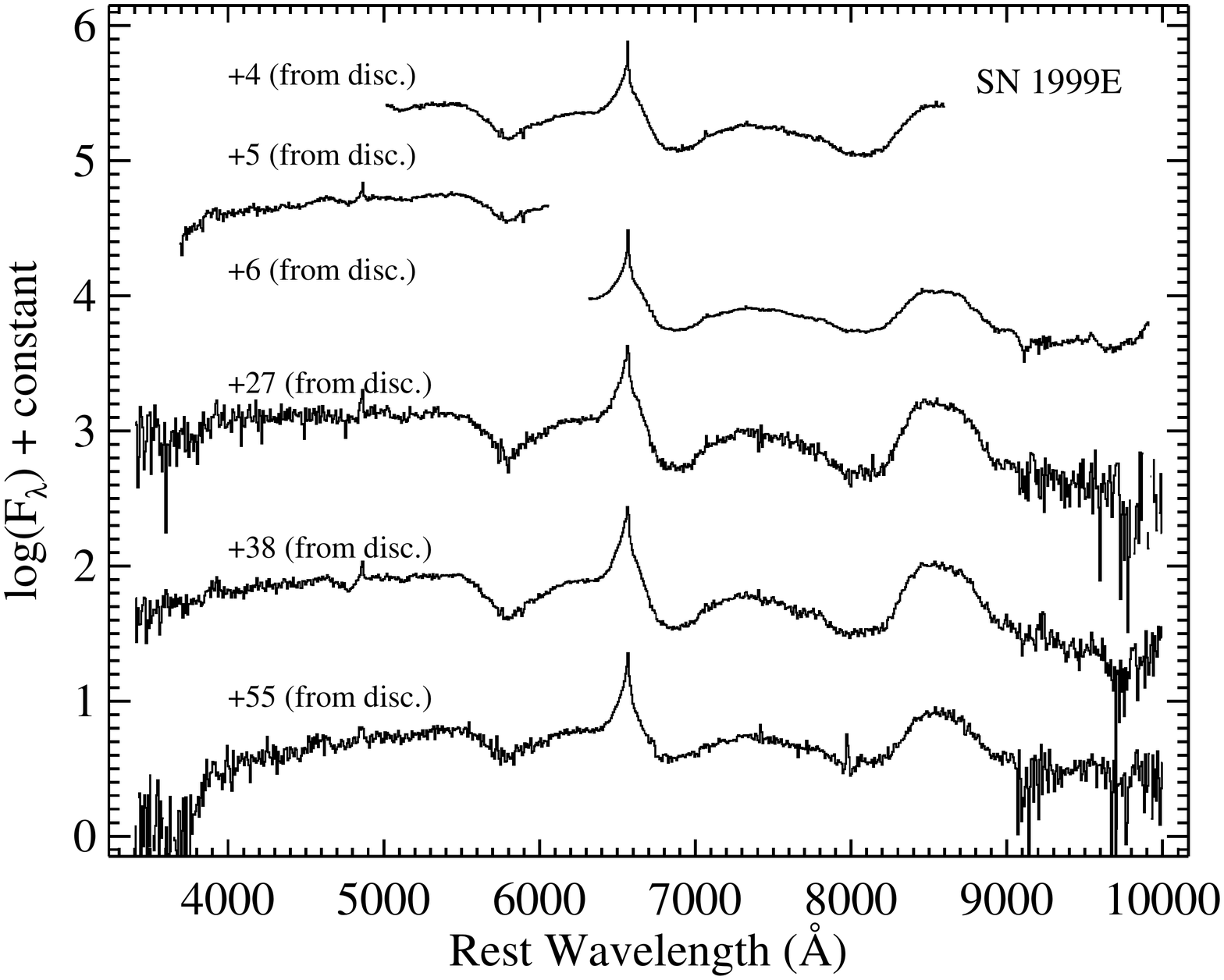}
\caption{Spectra of SN~1999E, originally published in
  \citet{Filippenko00}, labeled with age relative to the discovery
  date. The data have had their host-galaxy recession velocity  
  removed and have been corrected for Galactic
  reddening.}\label{f:sn1999e}
\end{figure*}

The two next-best-studied SNe~Ia-CSM are SN~1997cy
\citep{Turatto00,Germany00} and SN~1999E \citep{Filippenko00,Rigon03}. 
Neither of these objects have as much data as either
SNe~2002ic or 2005gj, but they have still been studied fairly
rigorously due to their peculiar nature (especially before the
discovery of SNe~2002ic and 2005gj). Again, both objects showed
relatively bright absolute magnitudes with spectra that somewhat
resembled those of SNe~Ia with superposed \halpha\ emission. 
In Figure~\ref{f:sn1999e} we
plot our six spectra of SN~1999E, which were first published by
\citet{Filippenko00}, before it was realized that this was a
SN~Ia. Unfortunately, it seems that SN~1999E was
discovered well after maximum brightness; thus, the phases displayed
in Figure~\ref{f:sn1999e} are relative to the UT date of discovery
\citep[1999~Jan.~15;][]{Rigon03}, and are likely from a much later
phase if calculated relative to maximum brightness.

\subsection{SN~2008J and SN~2008cg}\label{ss:08j_08cg}

\begin{figure*}
\centering
\includegraphics[width=6.8in]{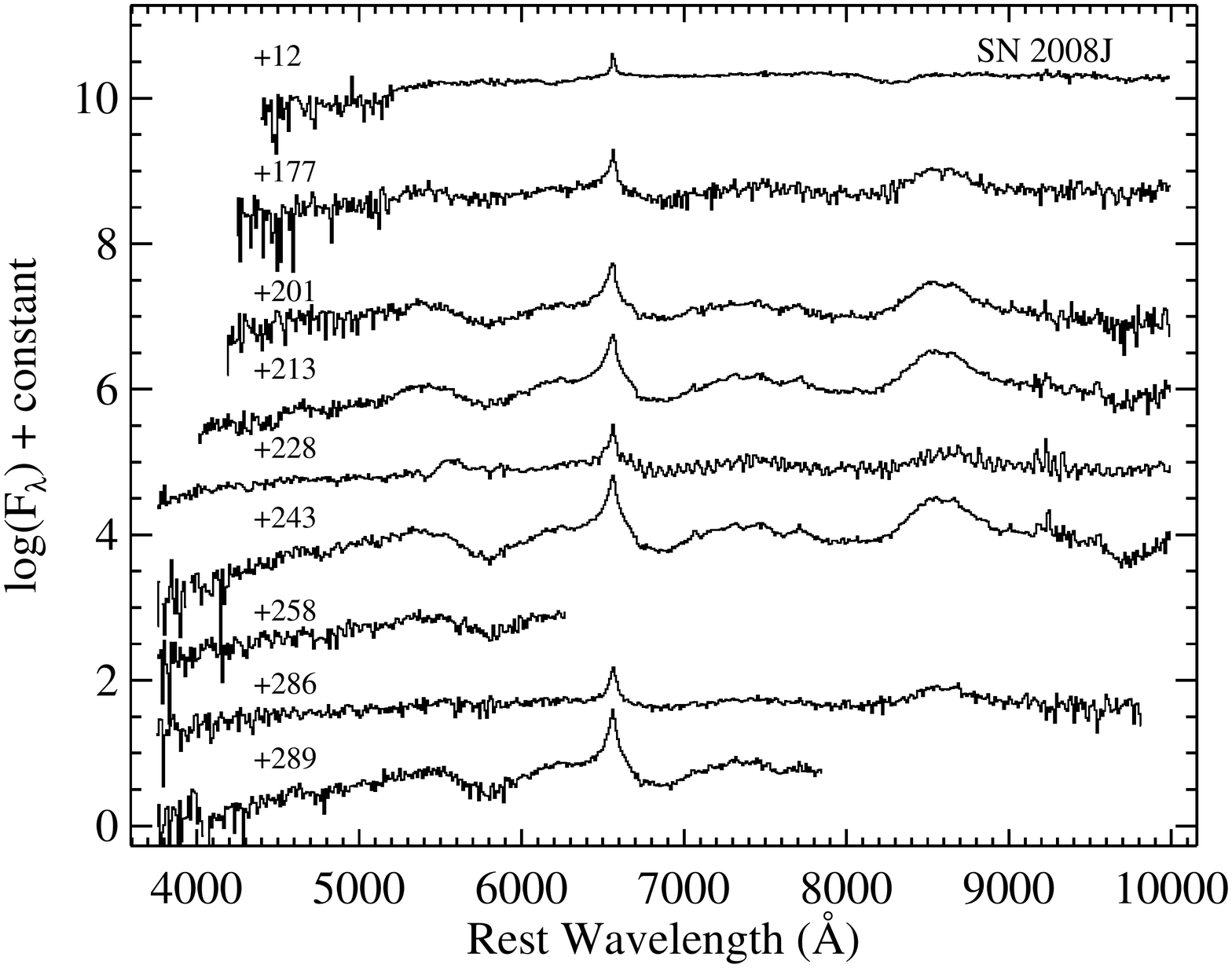}
\caption{Spectra of SN~2008J, the \halpha\ profiles of which were
  analyzed by \citet{Fox11}, labeled with age relative to maximum
  brightness. The data have had their host-galaxy recession velocity 
  removed and have been corrected for Galactic
  reddening.}\label{f:sn2008j}
\end{figure*}

\begin{figure*}
\centering
\includegraphics[width=6.8in]{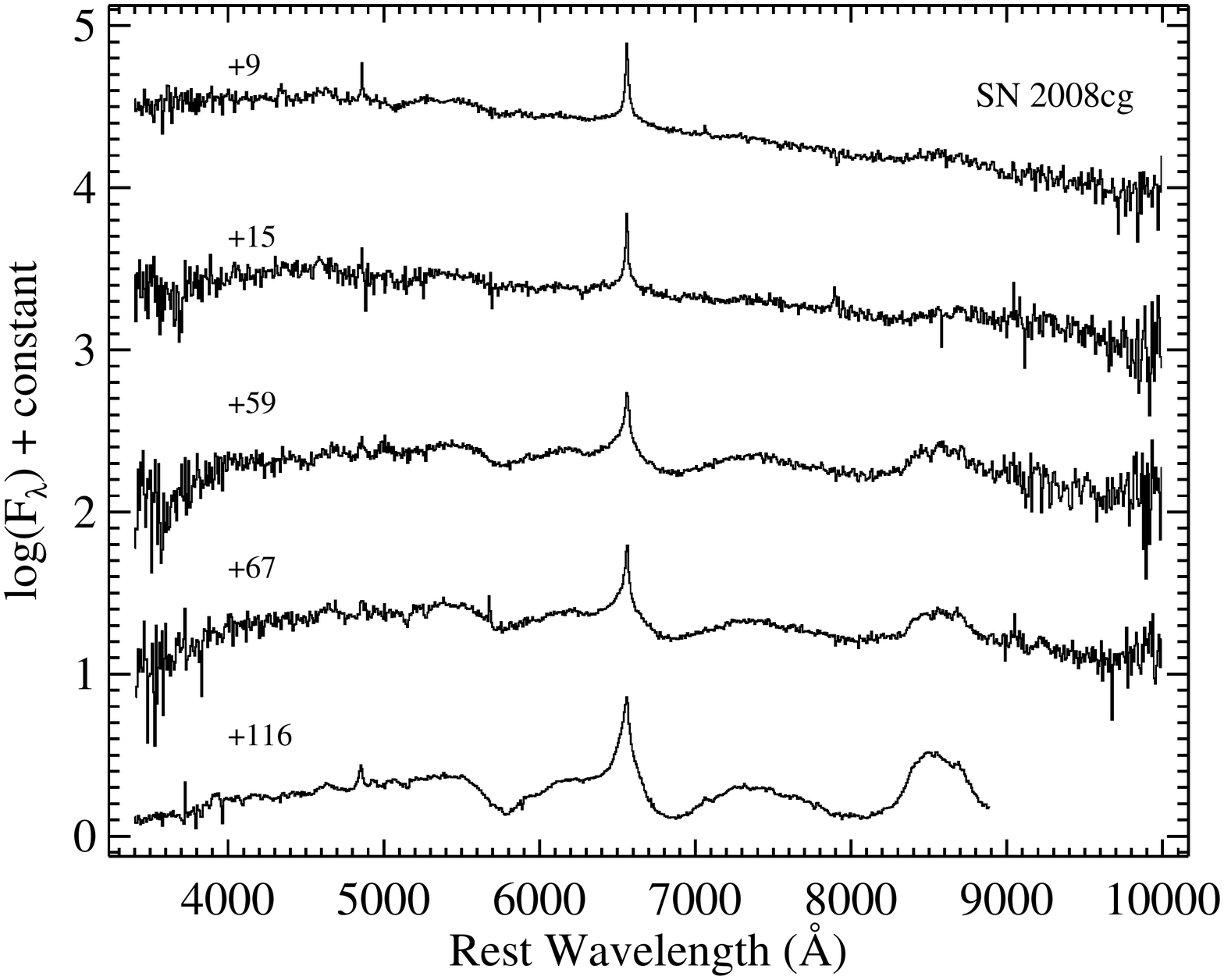}
\caption{Spectra of SN~2008cg, the \halpha\ profiles of which were
  analyzed by \citet{Fox11}, labeled with age relative to maximum
  brightness. The data have had their host-galaxy recession velocity 
  removed and have been corrected for Galactic
  reddening.}\label{f:sn2008cg}
\end{figure*}

SNe~2008J and 2008cg were both claimed to resemble SNe~2002ic and
2005gj in unrefereed classification circulars
\citep{08j_class,08cg_class2}, and very recently \citet{Taddia12} has
published an analysis of SN~2008J. \citet{Fox11} investigated the
\halpha\ profiles and presented mid-infrared (IR) photometry from a {\it
  Spitzer}/IRAC survey of these objects (referred to as
simply ``SNe~IIn'') and found that both SNe were clearly detected in
the mid-IR \about500--600~d after discovery (see
\S\ref{ss:other_observations} for more information). SN~2008J
resembles SN~2005gj in an optical spectrum obtained only 2~d after
discovery \citep{08j_class}, which is \about2.5~weeks before $r$-band
maximum brightness \citep{Taddia12}, and it continues to resemble
SNe~Ia-CSM in all nine of our spectra (plotted in
Figure~\ref{f:sn2008j}). \citet{Taddia12} also found that the optical
and NIR spectra of SN~2008J show narrow H Balmer, Paschen, and
Brackett emission (most with P-Cygni profiles) and that the optical
spectra could be decomposed into a low-order polynomial continuum and
the spectrum of the overluminous Type Ia SN~1991T. On the other hand,
an optical spectrum of SN~2008cg resembles a relatively normal SN~IIn
3~d after discovery \citep{08cg_class}, which is about 10~d past
maximum brightness. However, by \about2~m past discovery, SN~2008cg
was found  to closely resemble SNe~2002ic and 2005gj
\citep{08cg_class2}. Our five spectra of this objects are shown in
Figure~\ref{f:sn2008cg}. Note that the data on SNe~2008J and 2008cg
presented here are the same optical spectra discussed by
\citet{Fox11}.

\subsection{SN~2011jb and CSS120327:110520--015205}\label{ss:11jb_12c1}

\begin{figure*}
\centering
\includegraphics[width=6.8in]{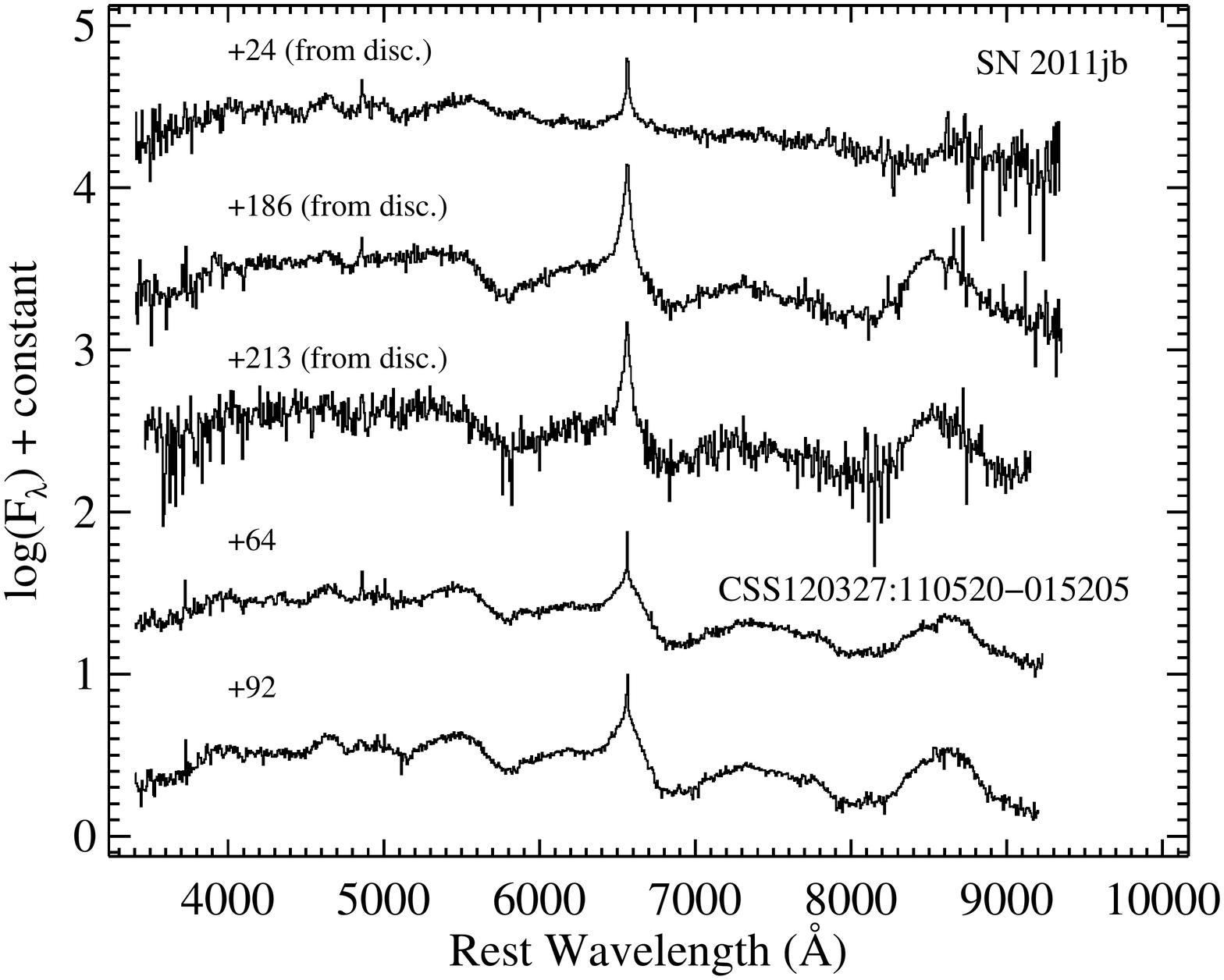}
\caption{Spectra of SN~2011jb labeled with age relative to the discovery
  date and spectra of CSS120327:110520--015205 labeled with age
  relative to maximum  
  brightness. The data have had their host-galaxy recession velocity 
  removed and have been corrected for Galactic
  reddening.}\label{f:sn2011jb}
\end{figure*}

Almost nothing has appeared in the literature regarding SNe~2011jb and
CSS120327:110520--015205 \citep{12c1_class},
despite both objects being immediately classified as SNe~Ia-CSM
\citep{11jb_class,12c1_class}. We present our three spectra of
SN~2011jb and our two spectra of CSS120327:110520--015205 in
Figure~\ref{f:sn2011jb}. 

\section{PTF Ia-CSM Objects}\label{s:ptf_iaCSM}

\begin{figure*}
\centering
\includegraphics[width=6.8in]{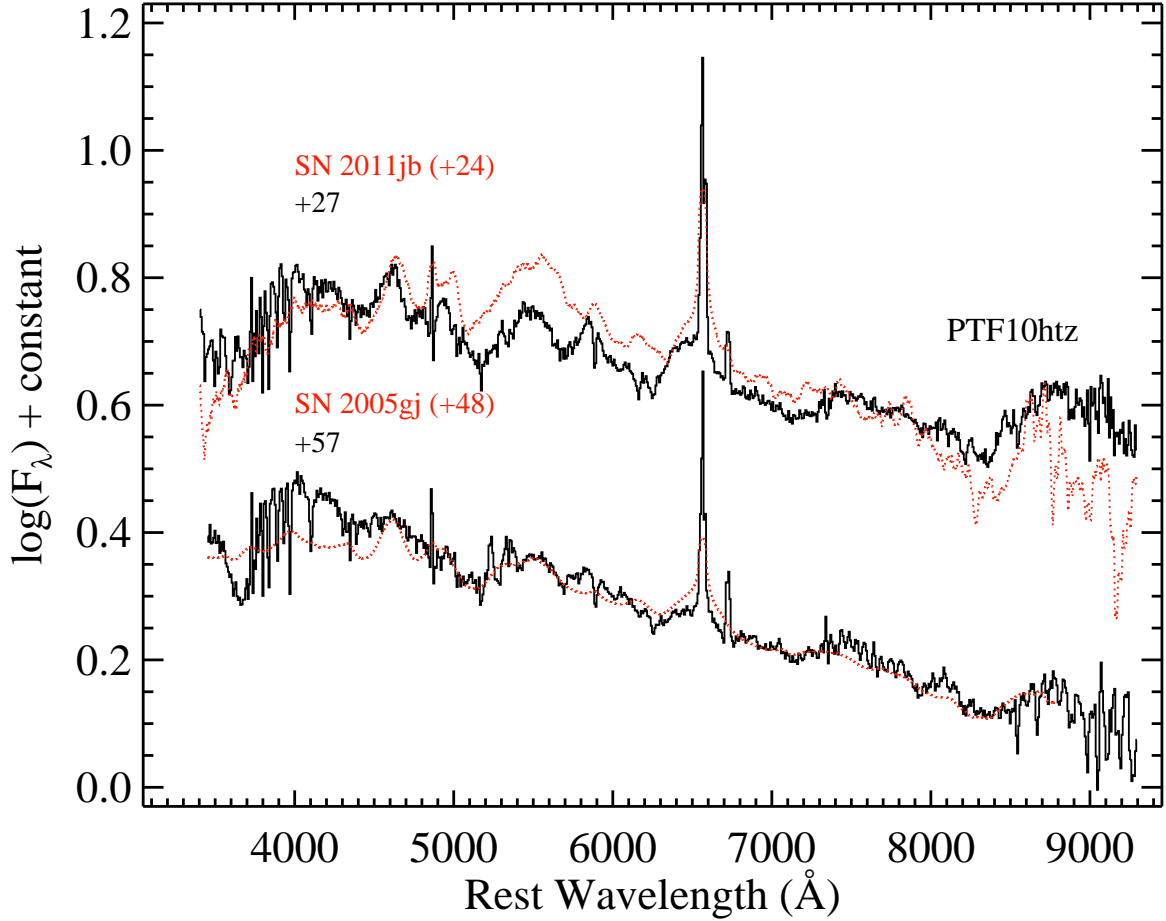}
\caption{Spectra of 
  PTF10htz labeled with age relative
  to maximum 
  brightness, and a comparison spectrum (red). The data have had their
  host-galaxy recession velocity 
  removed and have been corrected for Galactic
  reddening. Note that this object has a relatively large amount of
  host-galaxy contamination, but the superposed SN features match
  those of other SNe~Ia-CSM quite well.}\label{f:09hih} 
\end{figure*}

\begin{figure*}
\centering
\includegraphics[width=6.8in]{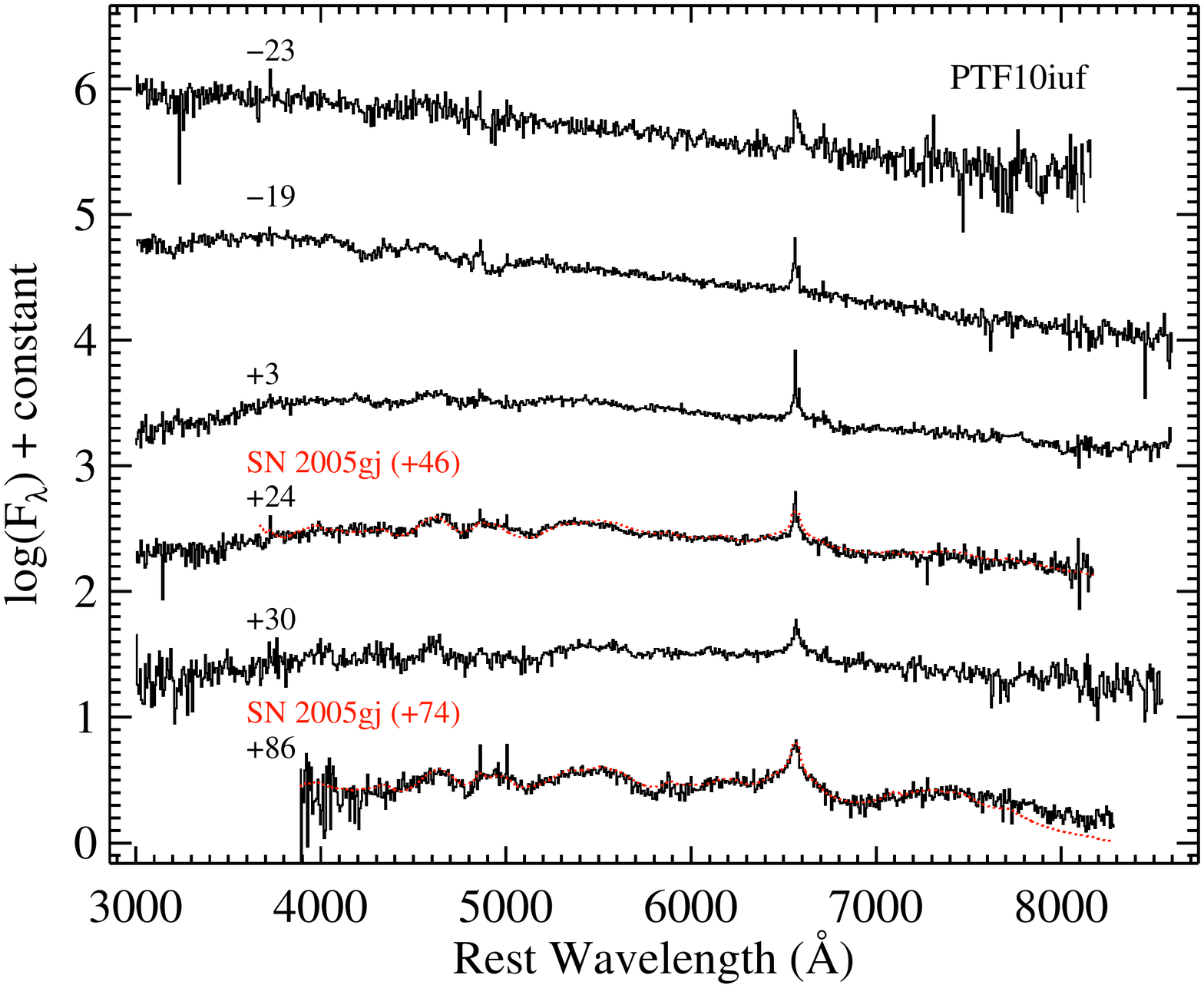}
\caption{Spectra of PTF10iuf labeled with age relative to maximum
  brightness, and some comparison spectra (red). The data have had
  their host-galaxy recession velocity 
  removed and have been corrected for Galactic
  reddening.}\label{f:10iuf} 
\end{figure*}

\begin{figure*}
\centering
\includegraphics[width=6.8in]{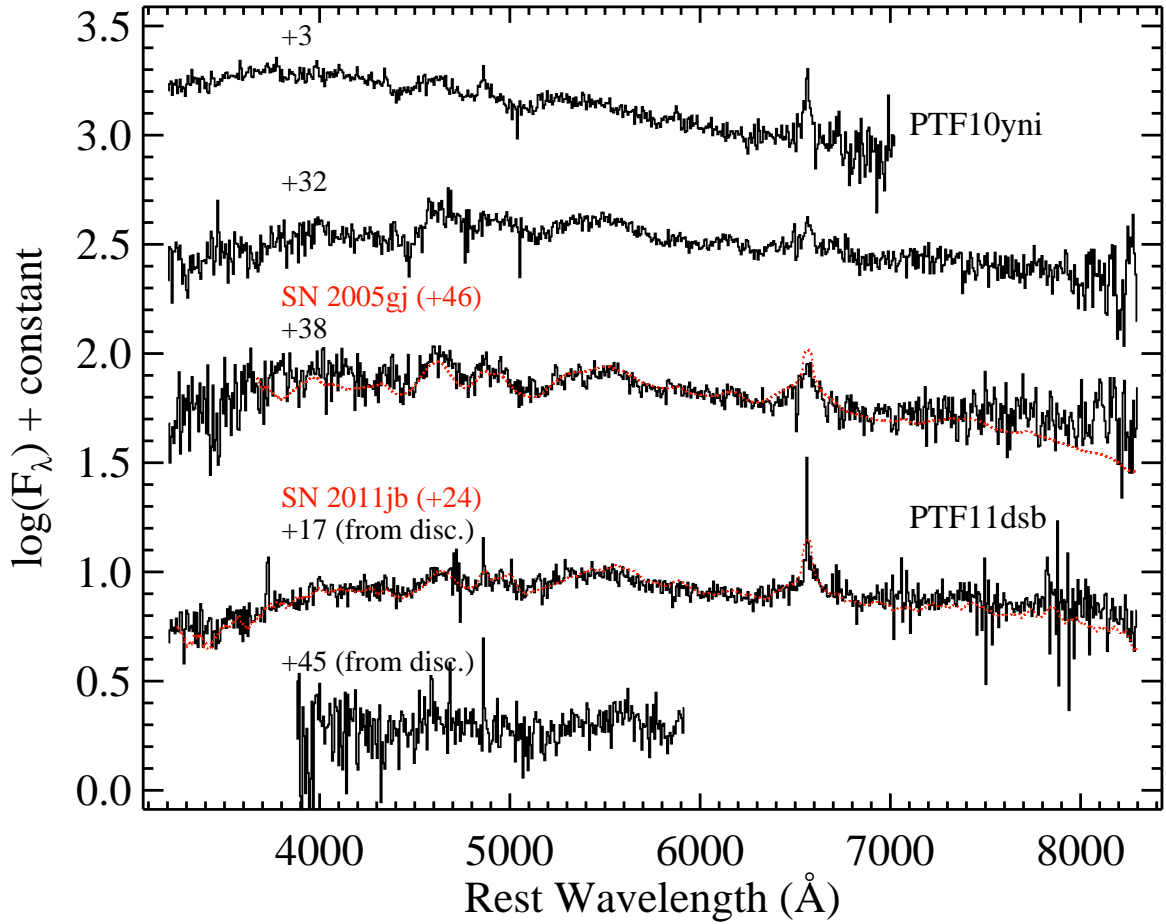}
\caption{Spectra of PTF10yni labeled with age relative to maximum
  brightness and spectra of PTF11dsb labeled with age relative to the
  discovery date, and some comparison spectra (red). The data have had their host-galaxy recession velocity
  removed and have been corrected for Galactic
  reddening.}\label{f:10yni} 
\end{figure*}

\begin{figure*}
\centering
\includegraphics[width=6.8in]{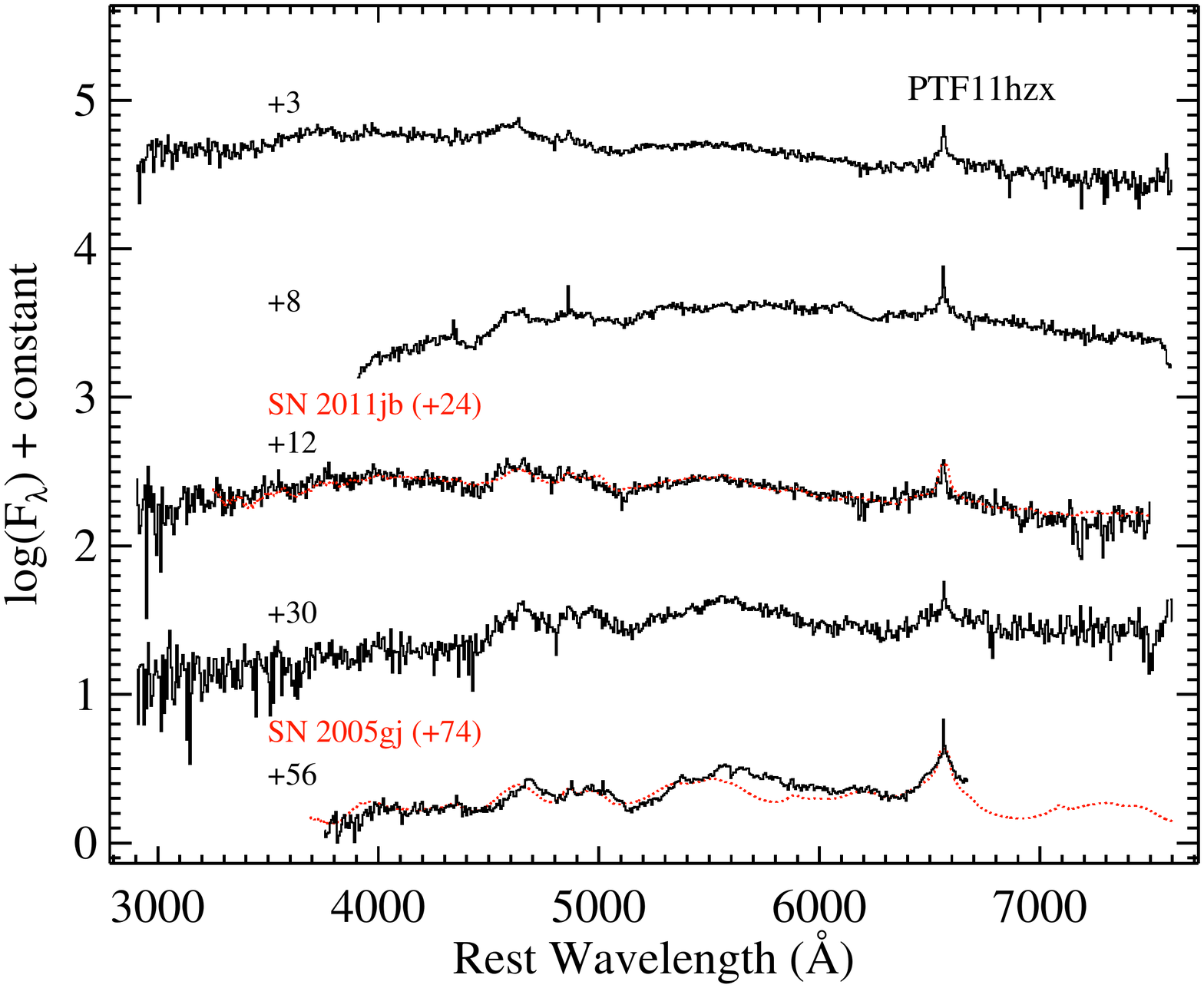}
\caption{Spectra of PTF11hzx labeled with age relative to maximum
  brightness, and some comparison spectra (red). The data have had their host-galaxy recession velocity
  removed and have been corrected for Galactic
  reddening.}\label{f:11hzx} 
\end{figure*}

\begin{figure*}
\centering
\includegraphics[width=6.8in]{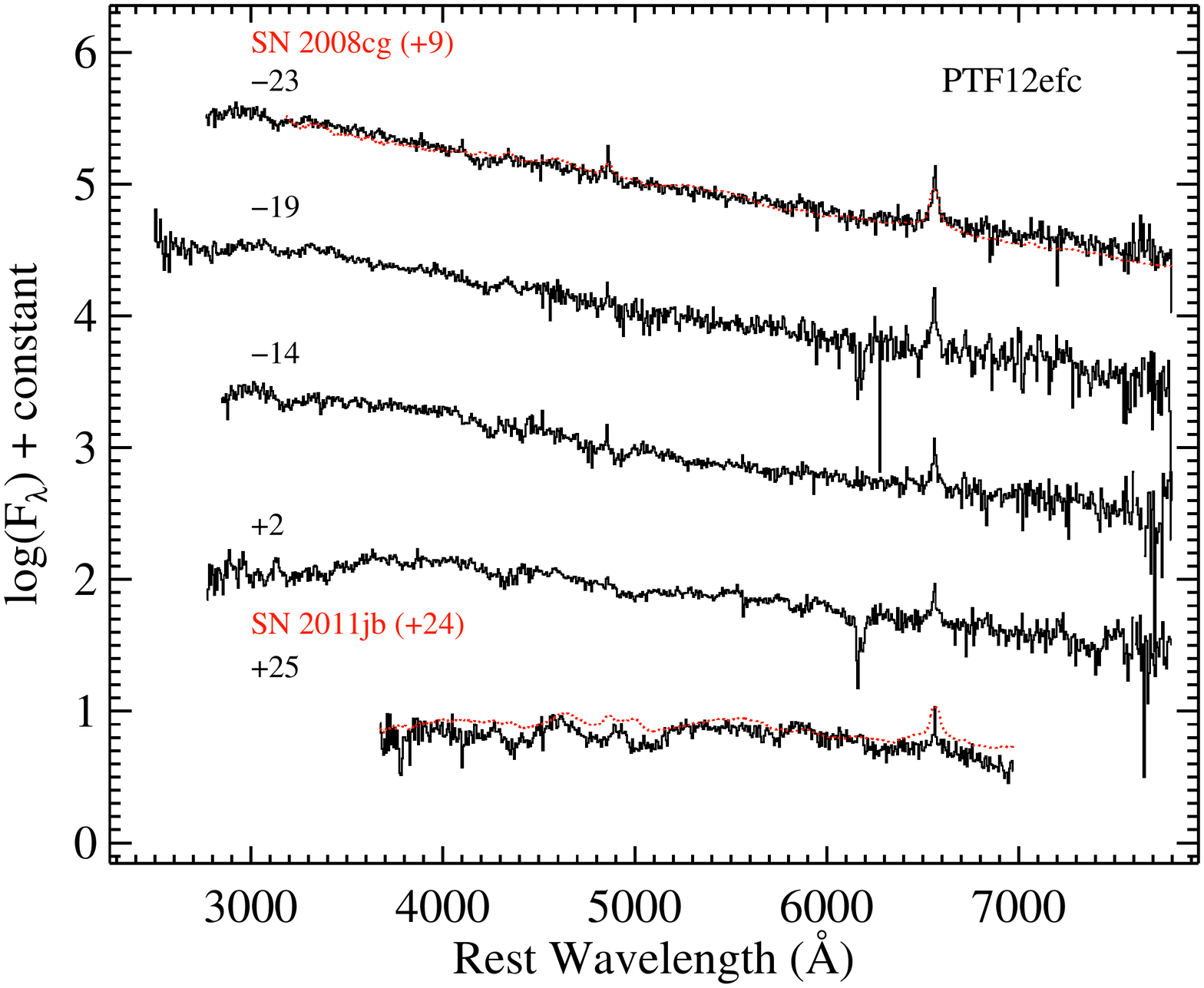}
\caption{Spectra of PTF12efc labeled with age relative to maximum
  brightness, and some comparison spectra (red). The data have had their host-galaxy recession velocity
  removed and have been corrected for Galactic
  reddening.}\label{f:12efc} 
\end{figure*}

\begin{figure*}
\centering
\includegraphics[width=6.8in]{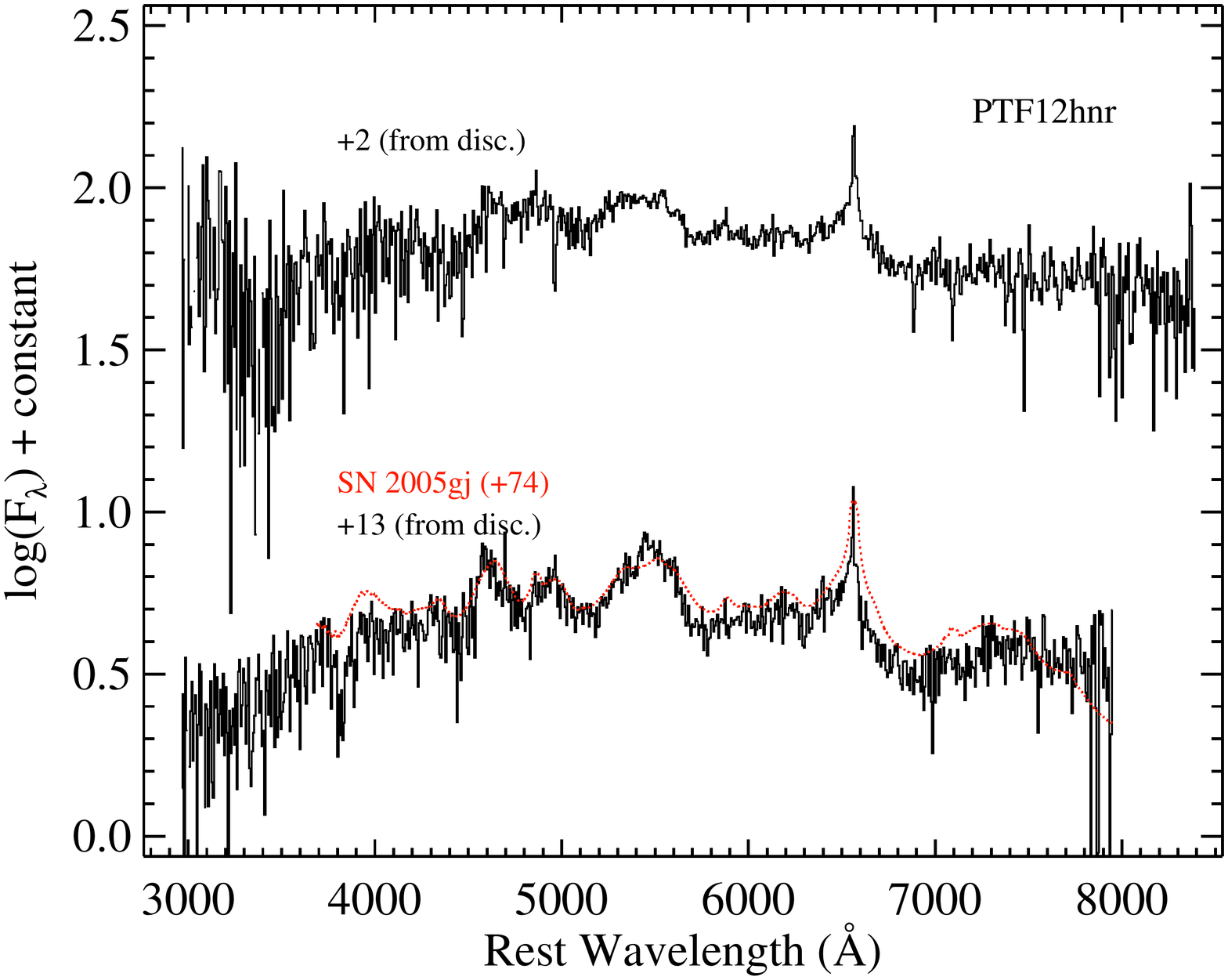}
\caption{Spectra of PTF12hnr labeled with age relative to the
  discovery date, and some comparison spectra (red). The data have had their host-galaxy recession velocity
  removed and have been corrected for Galactic
  reddening.}\label{f:12hnr} 
\end{figure*}

A similar search for SNe~Ia-CSM was performed using 178 spectra of
all 63 SNe~IIn discovered by PTF through August~2012. 
Once again, we ran these spectra through SNID to see if they were well
matched to our sample of SNe~Ia-CSM. This initial analysis yielded
nine possible objects, but two of these were rejected based on equally
good spectral matches to SNe~IIn. We classify the seven remaining
objects as true SNe~Ia-CSM. These PTF objects (in addition to
PTF11kx) and their host galaxies are summarized in
Tables~\ref{t:iaCSM_ptf} and \ref{t:iaCSM_ptf_host}, and our spectra of
them are described in Table~\ref{t:iaCSM_spec_ptf} and displayed in 
Figures~\ref{f:09hih}--\ref{f:12hnr}. See \citet{Dilday12} and
Silverman \etal (submitted) for further information regarding PTF11kx 
and its spectra.

\begin{table*}
\begin{center}
\caption{PTF SNe~Ia-CSM}\label{t:iaCSM_ptf}
\begin{tabular}{lcccc}
\hline
\hline
SN Name & Discovery & Approx. Date of & Peak Absolute & Galactic \\
                   & Date & $r$-Band Maximum & $r$-Band Magnitude & Reddening$^\textrm{a}$ \\
\hline
PTF11kx & 2011~Jan.~16 & 2011~Jan.~29 & $-19.3$ & 0.053 \\
\hline
PTF10htz & 2010~Apr.~3 & 2010~May~16 & $-19.1$ & 0.049 \\
PTF10iuf & 2010~June~5 & 2010~July~4 & $-20.5$ & 0.021 \\
PTF10yni & 2010~Oct.~3 & 2010~Oct.~30 & $-20.6$ & 0.059 \\
PTF11dsb & 2011~May~13 & $\cdots^\textrm{b}$ & $-19.8$ & 0.020 \\
PTF11hzx & 2011~July~17 & 2011~July~23 & $-21.3$ & 0.096 \\
PTF12efc & 2012~May~13 & 2012~June~15 & $-21.0$ & 0.015 \\
PTF12hnr & 2012~Aug.~7 & $\cdots^\textrm{b}$ & $-21.0$ & 0.037 \\
\hline\hline
\multicolumn{5}{p{5.9in}}{$^\textrm{a}$Galactic reddening toward each
  SN as derived from the dust maps of \citet{Schlegel98}; includes
  the corrections of \citet{Peek10}.} \\
\multicolumn{5}{p{5.9in}}{$^\textrm{b}$SN was discovered after maximum
  brightness.} \\
\hline\hline
\end{tabular}
\end{center}
\end{table*}

\begin{table*}
\begin{center}
\caption{PTF Ia-CSM Host Galaxies}\label{t:iaCSM_ptf_host}
\begin{tabular}{lccc}
\hline
\hline
SN Name &Name & Type & Redshift $z$ \\
\hline
\hline
PTF11kx & SDSS J080913.20+461842.9 & Late-Type Spiral & 0.0466 \\
\hline
PTF10htz & CGCG 352$-$058 & Late-Type Spiral & 0.0352 \\
PTF10iuf & SDSS J160615.65+335213.2 & Late-Type Spiral & 0.1586 \\
PTF10yni & $\cdots$ & $\cdots$ & 0.1688 \\
PTF11dsb & SDSS J161835.63+324150.0 & Late-Type Spiral & 0.1900 \\
PTF11hzx & $\cdots$ & $\cdots$ & 0.2287 \\
PTF12efc & $\cdots$ & $\cdots$ & 0.2341 \\
PTF12hnr & SDSS J232415.43$-$051239.0 & Late-Type Spiral & 0.1883 \\
\hline\hline
\end{tabular}
\end{center}
\end{table*}

\begin{table*}
\scriptsize
\begin{center}
\caption{Spectra of PTF SNe~Ia-CSM}\label{t:iaCSM_spec_ptf}
\begin{tabular}{lcccrr}
\hline
\hline
UT Date &  Age (d)$^\textrm{a}$  &  Instrument$^\textrm{b}$ & Range (\AA)  & Res. (\AA)$^\textrm{c}$ & Exp. (s) \\
\hline
\multicolumn{6}{c}{PTF10htz} \\
\hline
2010~June~13.3  & $\phantom{-}$27 & DBSP & 3500--9800$\phantom{1}$ & 3/4 & 450 \\
2010~July~14.2  & $\phantom{-}$57 & DBSP & 3550--9900$\phantom{1}$ & 3/4 & 450 \\
\hline
\multicolumn{6}{c}{PTF10iuf} \\
\hline
2010~June~7.1  & $-23$ & ISIS & 3150--9500$\phantom{1}$ & 3.5/7.2 & 900 \\
2010~June~12.5  & $-19$ & LRIS & 3258--10116 & 4.5/6.2 & 900 \\
2010~July~7.4  & $\phantom{0-}$3 & LRIS & 3300--10200 & 4.5/6 & 450 \\
2010~Aug.~1.0  & $\phantom{-}$24 & ISIS & 3150--9500$\phantom{1}$ & 3.5/7.2 & 900 \\
2010~Aug.~8.3  & $\phantom{-}$30 & DBSP & 3400--9935$\phantom{1}$ & 3/4 & 750 \\
2010~Oct.~12.2  & $\phantom{-}$86 & DEIMOS & 4480--9630$\phantom{1}$ & 3 & 750 \\
\hline
\multicolumn{6}{c}{PTF10yni} \\
\hline
2010~Nov.~3.2  & $\phantom{0-}$3 & RCS & 3330--8235$\phantom{1}$ & 5.9 & 1200 \\
2010~Dec.~6.2  & $\phantom{-}$32 & DBSP & 3505--10000 & 3/4 & 600 \\
2010~Dec.~13.2  & $\phantom{-}$38 & DBSP & 3500--9900$\phantom{1}$ & 3/4 & 1200 \\
\hline
\multicolumn{6}{c}{PTF11dsb} \\
\hline
2011~June~2.1  & $\phantom{-}$17 & LRIS & 3040--10240 & 6.5/6 & 450 \\
2011~July~5.4  & $\phantom{-}$45 & DEIMOS & 4592--7070$\phantom{1}$ & 3 & 600 \\
\hline
\multicolumn{6}{c}{PTF11hzx} \\
\hline
2011~July~26.4  & $\phantom{0-}$3 & DBSP & 3440--9800$\phantom{1}$ & 3/4 & 900 \\
2011~Aug.~1.5  & $\phantom{0-}$8 & DEIMOS & 4765--9633$\phantom{1}$ & 3 & 1000 \\
2011~Aug.~6.4  & $\phantom{-}$12 & DBSP & 3400--9250$\phantom{1}$ & 3/4 & 600 \\
2011~Aug.~28.3  & $\phantom{-}$30 & DBSP & 3330--10000 & 3/4 & 430 \\
2011~Sep.~29.4  & $\phantom{-}$56 & DEIMOS & 4583--8232$\phantom{1}$ & 3 & 1200 \\
\hline
\multicolumn{6}{c}{PTF12efc} \\
\hline
2012~May~17.5  & $-23$ & LRIS & 3382--10108 & 3.7/5.9 & 450 \\
2012~May~22.4  & $-19$ & LRIS & 3050--10257 & 6.5/6 & 480 \\
2012~May~29.3  & $-14$ & DBSP & 3480--10400 & 3/4 & 1200 \\
2012~June~18.3  & $\phantom{0-}$2 & DBSP & 3388--10120 & 3/4 & 1800 \\
2012~July~16.3 & $\phantom{-}$25 & DEIMOS & 4500--8640$\phantom{1}$ & 3 & 800 \\
\hline
\multicolumn{6}{c}{PTF12hnr} \\
\hline
2012~Aug.~9.5  & $\phantom{0-}$2 & Kast & 3500--10000 & 4/10 & 2100 \\
2012~Aug.~22.1 & $\phantom{-}$13 & ISIS & 3500--9476$\phantom{1}$ & 3.5/7.2 & 1800 \\
\hline\hline
\multicolumn{6}{p{4.2in}}{$^\textrm{a}$Rest-frame days relative to
  maximum brightness, except for PTF11dsb and PTF12hnr where the epoch 
  is relative to the date of discovery (2011~May~13 and 2012~Aug.~7,
  respectively). See 
  Table~\ref{t:iaCSM_ptf} for the dates of maximum 
  brightness for the rest of the objects.} \\
\multicolumn{6}{p{4.2in}}{$^\textrm{b}$DBSP = Double Spectrograph
  on the Palomar 200~inch telescope; ISIS = Intermediate dispersion
  Spectrograph and Imaging System on the 4.2~m William Herschel
  Telescope; LRIS = Low Resolution Imaging
  Spectrometer on the Keck 10~m telescope; DEIMOS = DEep Imaging
  Multi-Object Spectrograph on the 
  Keck 10~m telescope; RCS = RC Spec on the KPNO 4~m telescope; Kast =
  Kast double 
  spectrograph on the Shane 3~m telescope at Lick Observatory.} \\
\multicolumn{6}{p{4.2in}}{$^\textrm{c}$Approximate FWHM resolution. If
  two numbers are listed, 
  they represent the blue-side and red-side resolutions, respectively.} \\ 
\hline\hline
\end{tabular}
\end{center}
\end{table*}
\normalsize

Three of the PTF SNe~Ia-CSM were publicly announced while the
other four were not. PTF10htz was initially classified as a Type IIb
SN \citep{10htz_class}, but our careful SNID analysis shows that it is
likely a SN~Ia-CSM (with substantial host-galaxy
contamination) and matches SNe~2011jb and 2005gj
(Fig.~\ref{f:09hih}). In \citet{10yni_class}, 
PTF10yni \citep[which was independently discovered by][and called
CSS101008:001049+141039]{10yni_disc} was classified as a ``IIn/Ic,''
but once again our SNID analysis indicates that it is actually a
SN~Ia-CSM, matching SN~2005gj (Fig.~\ref{f:10yni}). Finally, PTF11dsb 
was classified simply as a ``SN II'' by \citet{11dsb_class}, but our
analysis here shows that it is possibly a SN~Ia-CSM, somewhat similar to
SN~2011jb (Fig.~\ref{f:10yni}), though it is not completely clear
whether this object is a {\it bona fide} member of the class.


\section{Analysis and Discussion of the SN~Ia-CSM Class}\label{s:analysis}

Using our SNID analysis we have identified seven new SNe~Ia-CSM from PTF 
(\S\ref{s:ptf_iaCSM}) and reidentified eight previously known 
SNe~Ia-CSM (\S\ref{s:iaCSM}). Given these 15 objects, in addition to
PTF11kx, we now attempt to define observational characteristics
of this class of SNe~Ia (while keeping in mind that PTF11kx may be an
extreme member). Though no single observable appears to be a sufficient
condition for a SN to be considered a Ia-CSM object, there seem to be a
handful of features that nearly all of the SNe~Ia-CSM
display. These are discussed in detail below, along with a few
observables that are also found more generally in ``normal'' SNe~IIn.

\subsection{Optical Photometry of SNe~Ia-CSM}\label{ss:phot}

\begin{figure*}
\centering
\includegraphics[width=6.8in]{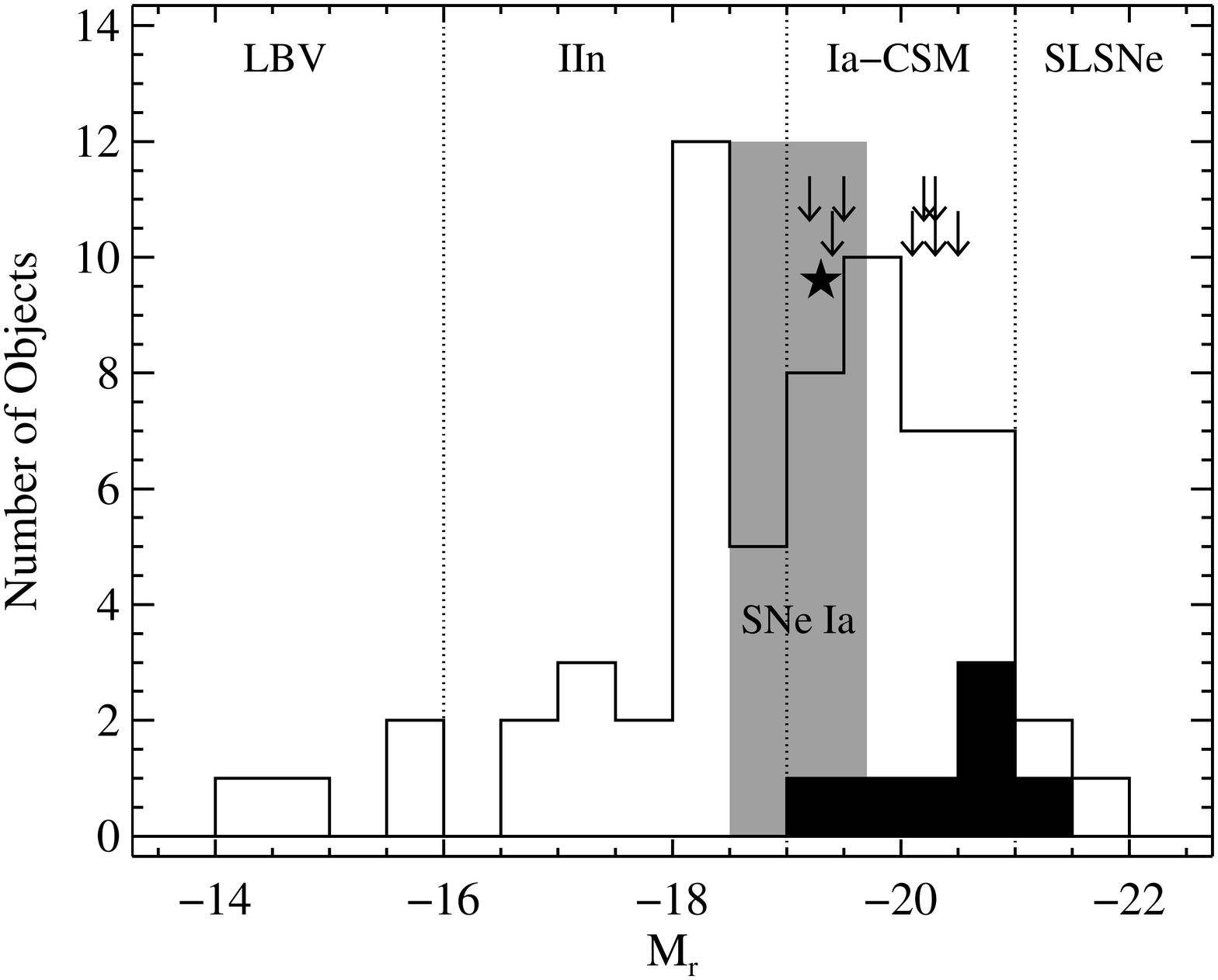}
\caption{Peak absolute $r$-band magnitude of all 63 SNe~IIn discovered
  by PTF through August~2012. The vertical dotted lines denote the
  boundaries between the various subtypes of SNe~IIn and SNe~Ia-CSM, 
  and the gray shaded region is the range of SNe~Ia that follow the
  Phillips relation. The black, filled histogram shows the seven SNe~Ia-CSM
  discovered by PTF (not including PTF11kx), the downward-pointing
  arrows are the eight previously known SNe~Ia-CSM, and the star is
  PTF11kx.}\label{f:ptf_phot}
\end{figure*}

The Caltech Core-Collapse Project \citep[CCCP;][]{Kiewe12} found that
the typical peak absolute magnitude range for SNe~IIn is $-18.7 \leq
M_R \leq -17$~mag (using $H_0 = 73$~\kms~Mpc$^{-1}$, the
value adopted throughout this work), while the Lick Observatory
SN Search (LOSS) found their SNe~IIn to have peak luminosities in the
range $-19 \leq M_R \leq -16$~mag \citep{Li11a}. LOSS also showed that
SNe~IIP/IIL have an overlapping range of peak absolute magnitudes
\citep[$-17 \leq M_R \leq -15$~mag;][]{Smith11:LBV}, but the spectra
of SNe~IIP/IIL differ significantly from those of SNe~IIn or SNe~Ia-CSM 
\citep[the former consisting of broad P-Cygni profiles of H
Balmer lines; e.g.,][]{Filippenko97}. On the less luminous end,
luminous blue variable star (LBV) outbursts and so-called ``SN
impostors'' (which are spectroscopically similar to SNe~IIn and 
SNe~Ia-CSM) have peak absolute magnitudes $-16 \leq M_R$~mag
\citep{Smith11:LBV}. On the more luminous end, Type II superluminous
SNe (which can also resemble SNe~IIn and SNe~Ia-CSM
spectroscopically) have peak luminosities $M_R \la -21$~mag
\citep{Gal-Yam12,Quimby13}. We note that both CCCP and LOSS have a
dearth of objects which are spectroscopically similar to SNe~IIn with
peak absolute magnitudes $-21 \leq M_R \leq -19$~mag, although the PTF
sample contains quite a few objects in this range
(Fig.~\ref{f:ptf_phot}). 

Extending this analysis, in Figure~\ref{f:ptf_phot} we plot the peak
absolute $r$-band magnitude of all 63 objects spectroscopically
identified as SNe~IIn by PTF through August~2012. The photometric
calibration of the PTF data is 
based on the Sloan Digital Sky Survey (SDSS) data when possible,
otherwise the PTF calibration and natural magnitude system are used
\citep{Ofek12a,Ofek12b}. The vertical dotted lines denote the
boundaries between the various subtypes of SNe~IIn (LBV outbursts and 
SN impostors, typical SNe~IIn, and superluminous SNe~II) and our proposed
range of SN~Ia-CSM luminosities. The gray shaded region is the range of
SNe~Ia that follow the Phillips relation, about $-19.7$ to $-18.5$~mag 
\citep[e.g.,][]{Ganeshalingam10:phot_paper}. The black, filled histogram
shows the peak absolute magnitudes of the seven SNe~Ia-CSM discovered by PTF
(\S\ref{s:ptf_iaCSM}) not including PTF11kx, the downward-pointing
arrows signify the peak luminosities of the eight previously known 
SNe~Ia-CSM (\S\ref{s:iaCSM}), and the star represents the peak 
absolute magnitude of PTF11kx \citep{Dilday12}.

As seen in Figure~\ref{f:ptf_phot}, 15 of the 16 SNe~Ia-CSM discussed in
this paper fall in the luminosity range $-21 \leq M_R \leq
-19$~mag. The one outlier, PTF11hzx, just barely misses this cutoff
with peak luminosity $M_R = -21.3$~mag. It appears that SNe~Ia-CSM 
{\it must} have an absolute optical magnitude in this range, roughly
0.5--1.5~mag brighter than their more normal SN~Ia cousins, likely
due to the interaction of the SN ejecta with the CSM (which gives
SNe~Ia-CSM their distinct spectra).


We also investigate the rise times of SN~Ia-CSM light curves (i.e., the
time elapsed between explosion and maximum
brightness). \citet{Ganeshalingam11} found that normal SNe~Ia have a
rise time of \about18~d in the $B$ band and \about20~d in the $V$ band
(after correcting for light-curve shape). Longer rise times have been
seen in more exotic SN~Ia subtypes, with possible super-Chandrasekhar
mass SNe~Ia having rise times of \about24~d
\citep[e.g.,][]{Scalzo10,Silverman11}.

Accurate rise-time measurements require photometric observations well
before maximum brightness; thus, not many of the previously studied
SNe~Ia-CSM have reasonable rise-time constraints. SN~2002ic was
found to have a rise time as long as 28~d \citep{Wood-Vasey04}, while
SN~2005gj had a rise time of \about20~d in the $g$ band and possibly up to
\about32~d in the $r$ band \citep{Aldering06,Prieto07}. On the other
hand, CSS120327:110520--015205 appeared to have a rise time of
\about45~d \citep{12c1_class}. As for the PTF SNe~Ia-CSM,
\citet{Dilday12} used a rise time of \about20~d for PTF11kx. Five of
the seven other SNe~Ia-CSM from PTF have constraining pre-maximum
brightness photometry and show evidence for relatively long rise times 
(\about30--40~d).

Thus, it appears that SNe~Ia-CSM tend to have longer rise times than more
normal SNe~Ia. This is consistent with the idea that the rise time is
related to the photon diffusion time (which should be longer in 
SNe~Ia-CSM as the light must make its way out of the relatively large
amount of CSM) as well as models of circumstellar shells in symbiotic
recurrent nova systems \citep{Moore12}. These rise times can also be
used to estimate the mass-loss rate of the progenitor system that gave
rise to the CSM. Using typical SN~Ia-CSM ejecta and wind velocities of a
few thousand and 100~\kms, respectively, and equations in \citet{Ofek13},
we find that SNe~Ia-CSM have mass-loss rates of a few times
$10^{-1}$~\msun~yr$^{-1}$.

\subsection{Optical Spectroscopy of SNe~Ia-CSM}\label{ss:spec}

We identified SNe~Ia-CSM using our SNID analysis outlined in
\S \ref{s:iaCSM} and \S \ref{s:ptf_iaCSM}. These matches were based solely
on comparing low-resolution optical spectra of input SNe with a
library of template spectra. When comparing the spectra of the SNe~Ia-CSM, 
some show relatively strong underlying SN~Ia features (e.g.,
SNe~2002ic and 2005gj, and especially PTF11kx), mostly at early times
and often resembling the somewhat overluminous Type Ia SN~1999aa
\citep{Li01:pec,Strolger02,Garavini04}. However, other SNe~Ia-CSM
(mostly ones with spectra from later epochs) do not resemble any
subtype of SN~Ia, aside from the other SNe~Ia-CSM, and are much
more easily mistaken for more typical SNe~IIn.

\begin{figure*}
\centering
\includegraphics[width=6.8in]{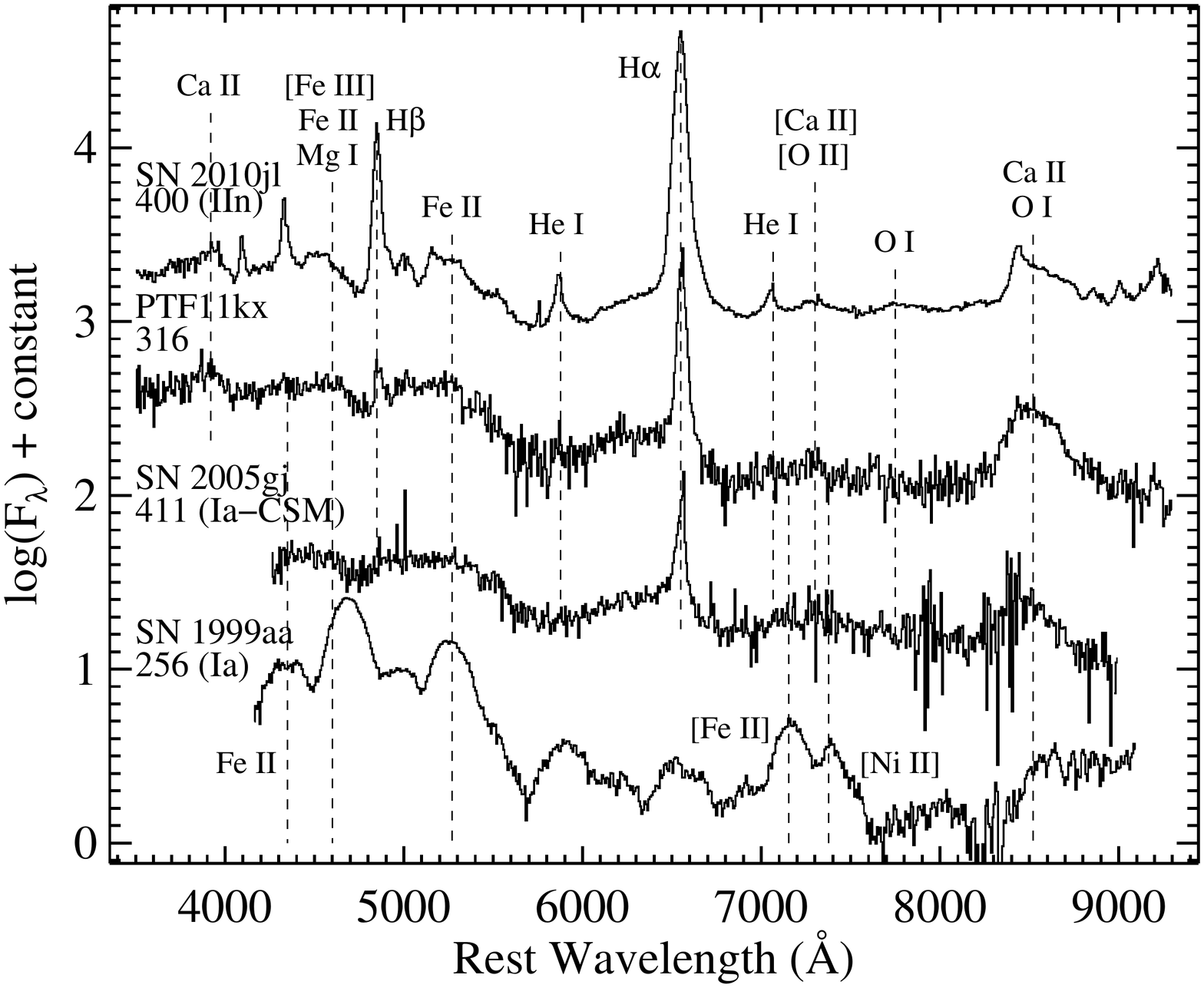}
\caption{Spectra of PTF11kx and SN~2005gj, in addition to SN~2010jl
  \citep[a SN~IIn;][]{Smith11} and SN~1999aa \citep[a somewhat
  overluminous SN~Ia;][]{Silverman12:BSNIPI}. Each spectrum is labeled
  with its rest-frame age relative to maximum brightness and major
  spectral features are labeled. The data have had their host-galaxy
  recession velocity removed and have been corrected for Galactic
  reddening. The figure is reproduced from Silverman \etal
  (submitted).}\label{f:11kx_comp}
\end{figure*}

At these late times, the spectra of SNe~Ia-CSM basically consist of a
relatively blue ``quasi-continuum,'' which is likely due to emission
from many overlapping, relatively narrow lines of iron-group elements
(IGEs), mostly \ion{Fe}{2}, excited by the CSM interaction, in
addition to strong \halpha\ emission and often broad \ion{Ca}{2}
emission \citep[e.g.,][]{Deng04}. Figure~\ref{f:11kx_comp} shows
spectra of PTF11kx and SN~2005gj, along with two comparison objects:
SN~2010jl \citep[a SN~IIn;][]{Smith11}, and SN~1999aa \citep[a
somewhat overluminous SN~Ia;][]{Silverman12:BSNIPI}. The figure is
reproduced from Silverman \etal (submitted) and some of the major 
spectral features are labeled. 

We do not find strong evidence of oxygen in SNe~Ia-CSM, which is often 
prominent in SNe~IIn in the form of the \ion{O}{1} $\lambda$7774 
feature. However, it is possible that part of the very broad emission
feature around 7400~\AA\ (see especially
Figures~\ref{f:sn2005gj}--\ref{f:sn2011jb}) is produced  
by \ion{O}{1} $\lambda$7774 blended with [\ion{O}{2}] $\lambda\lambda$7319, 
7330, along with the almost certain [\ion{Ca}{2}] $\lambda\lambda$7291, 
7324. Moreover, note that the broad emission just blueward of \halpha\ 
often seen in SNe~Ia-CSM might be due to [\ion{O}{1}] 
$\lambda\lambda$6300, 6364, adding to the evidence for oxygen
emission. However, it could more likely be related to the broad \halpha\ 
emission underlying the narrower components discussed at length below.

To further investigate the spectral
characteristics of SNe~Ia-CSM (specifically as compared to
SNe~IIn), we measure properties of the \halpha, \hbeta, and
\ion{He}{1} 
$\lambda$5876 features in all of the objects classified as SNe~IIn
from the Berkeley SN Group's database \citep{Silverman12:BSNIPI} as
well as the 63 SNe~IIn discovered by PTF through August~2012. We
follow the procedure of \citet{Dilday12} and fit the spectral features
with Gaussian profiles, using multiple components when necessary. 
In the following discussion, we are referring to the broader component
with FWHM of about 500--2000~\kms, unless otherwise
specified. Furthermore, we found that the similarities and differences
between SNe~IIn and SNe~Ia-CSM were relatively persistent at all
epochs, and thus in the analysis below we consider data on {\it all}
SNe~IIn and SNe~Ia-CSM at {\it all} epochs.

We find that SNe~Ia-CSM tend to have \halpha\ profiles which,
for the most part, match the bulk of the SN~IIn distribution. The
velocity associated with the peak of the \halpha\ emission in
SNe~Ia-CSM is slightly redshifted from the systemic velocity of
their host galaxies, whereas the SN~IIn peaks match their host-galaxy
velocities more closely. However, this redshift is not statistically
significant. Also, the equivalent width (EW) and FWHM of \halpha\ are
slightly smaller in SNe~Ia-CSM as compared with SNe~IIn, but
again this is not a statistically significant result. For SNe~Ia-CSM
having multiple spectra, we find that the EW of \halpha\ shows
strong fluctuations until \about100--150~d past maximum brightness,
perhaps indicating the SN ejecta are interacting with multiple CSM
shells of varying sizes and/or densities. At later epochs, the EW
tends to increase with time until \about1~yr past maximum, when the EW
possibly begins to decrease. This overall behavior has been seen
previously in PTF11kx (\citealt{Dilday12}; Silverman \etal
submitted), SN~2002ic \citep{Wang04}, and SN~2005gj
\citep{Aldering06,Prieto07}.  

The typical luminosities of \halpha\ emission in nearly all of the
SNe~Ia-CSM studied herein (including PTF11kx) fall in the range
(1--9)$ \times 10^{40}$~\ergps. The only object that lies outside
this range is 
CSS120327:110520--015205 (which has a luminosity of \about$3.9 \times  
10^{41}$~\ergps). Previously calculated \halpha\ luminosities of
SNe~Ia-CSM are consistent with this range. Specifically, SN~2002ic
had a luminosity of \about$5 \times 10^{40}$~\ergps\ \citep{Kotak04},
SN~2005gj had luminosities of (1--10)$ \times 10^{40}$~\ergps\
\citep{Aldering06,Prieto07}, and SN~2008J had a luminosity of
\about$1.3 \times 10^{41}$~\ergps\ \citep{Taddia12}.

In their respective references (and assuming shock velocities of a few
thousand \kms\ and wind velocities of \about100~\kms), the \halpha\ 
luminosities of these three SNe~Ia-CSM were converted into
mass-loss rates in the range (2--120)$ \times
10^{-4}$~\msun~yr$^{-1}$, which is quite a bit lower than the
mass-loss rates we infer using the rise times of SNe~Ia-CSM. This
may be due to the fact that these estimates usually assume that the
\halpha\ emission is produced by optically thin CSM being ionized by the SN
radiation field, whereas we will show below that the \halpha\ feature
is instead likely produced largely by collisional excitation.

\begin{figure*}
\centering
\centering$
\begin{array}{cc}
\includegraphics[width=3in]{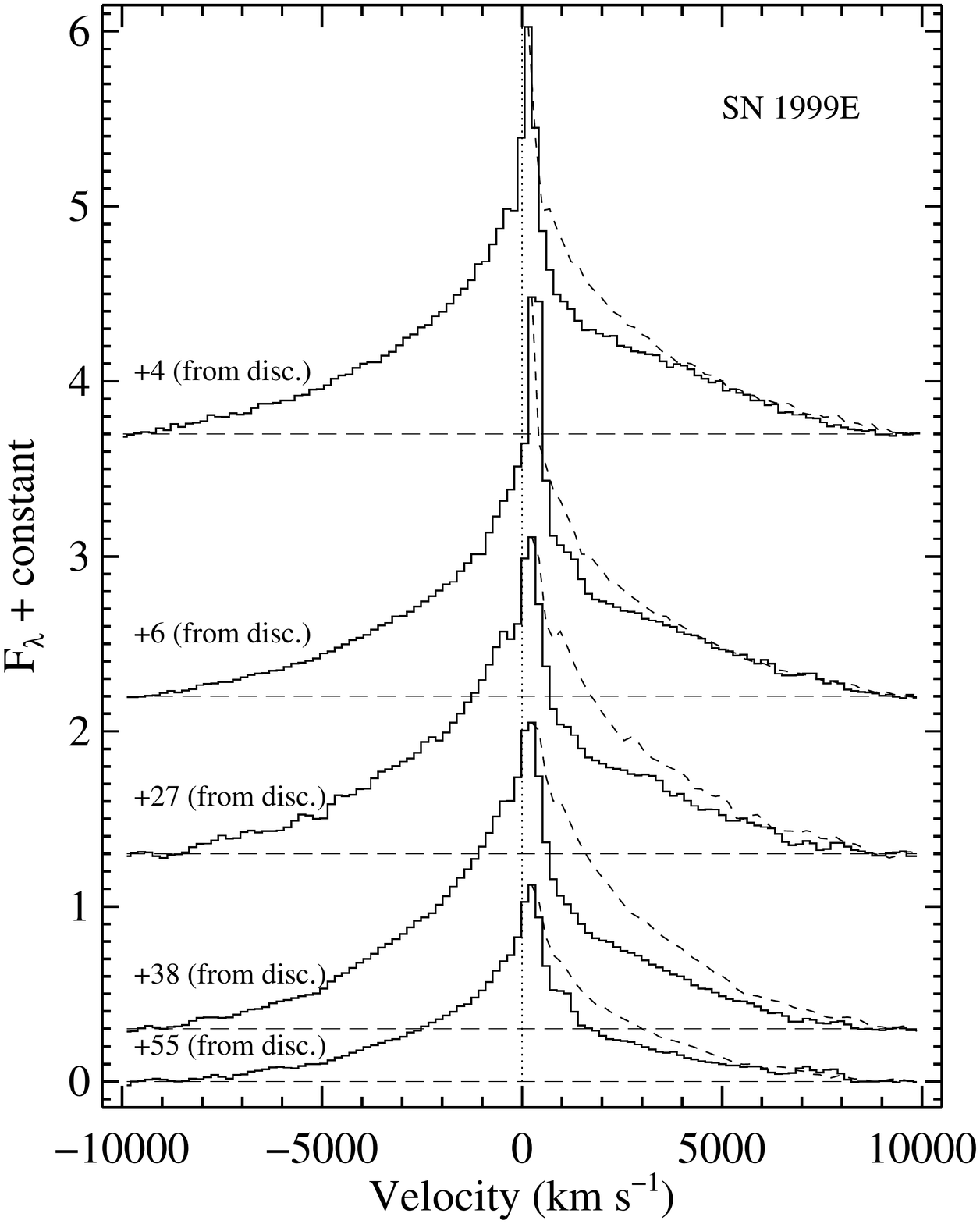} &
\includegraphics[width=3in]{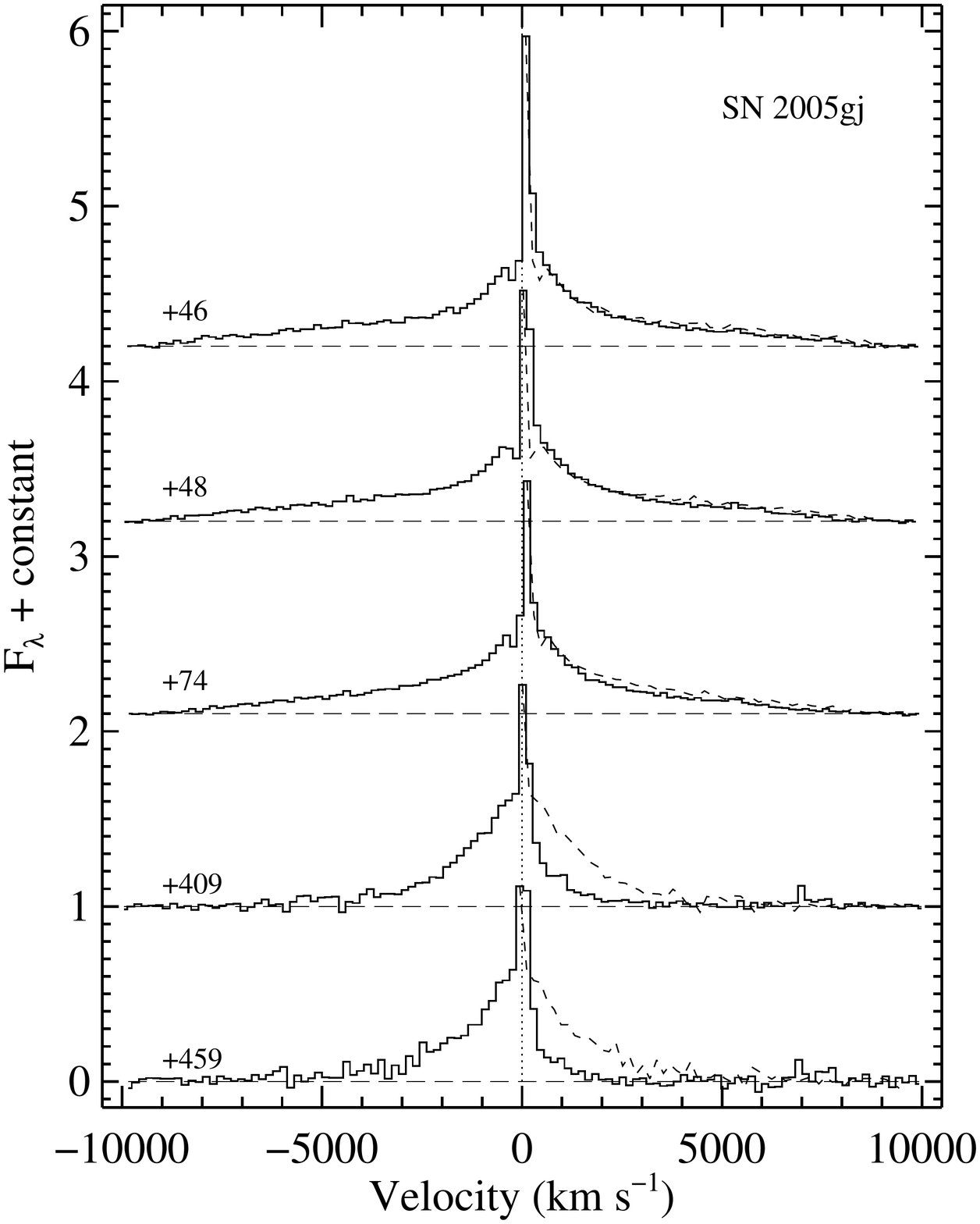} \\
\includegraphics[width=3in]{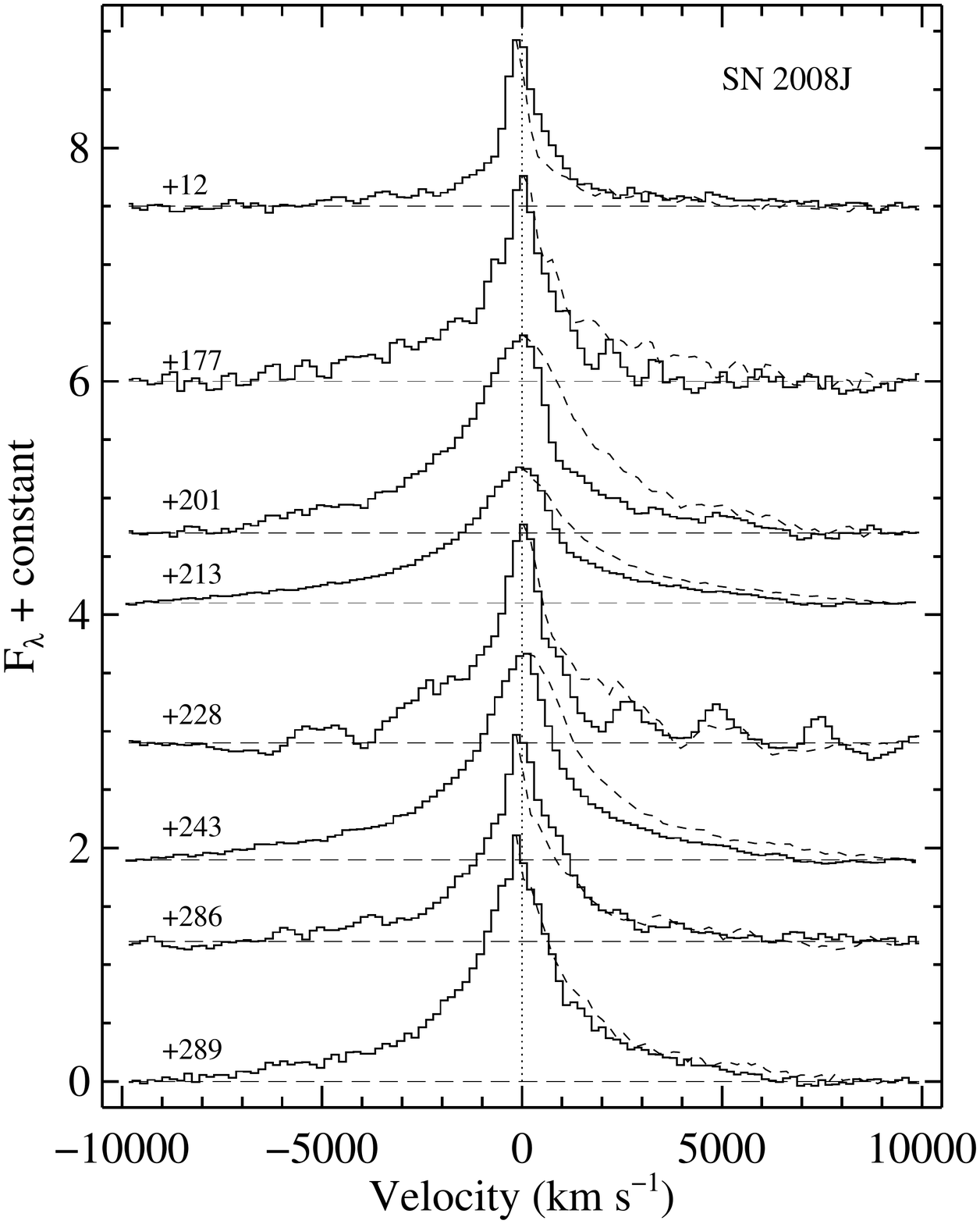} &
\includegraphics[width=3in]{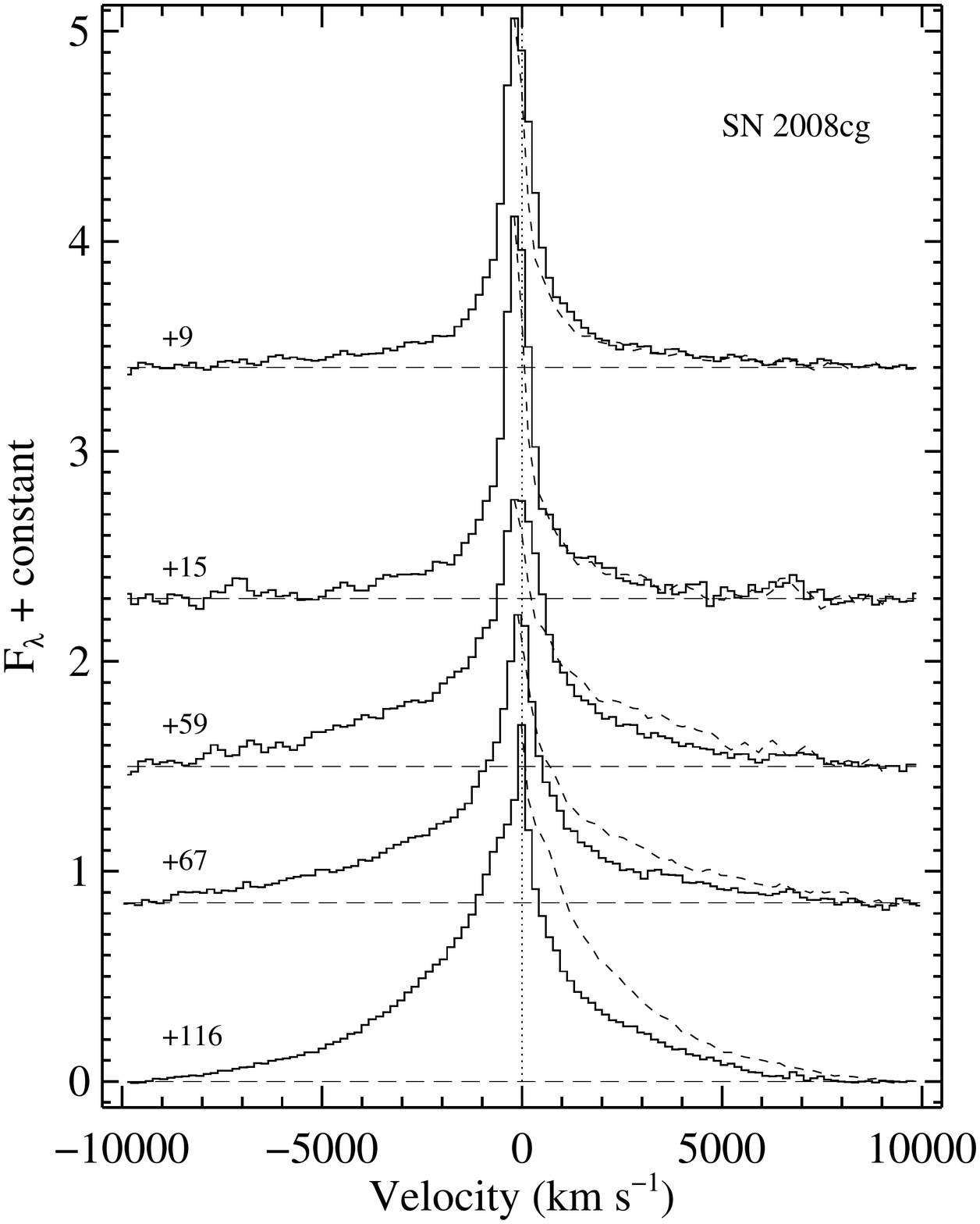} \\
\end{array}$
\caption{The \halpha\ profiles of SNe~1999E, 2005gj, 2008J, and
  2008cg. After removing a linear continuum (the long-dashed lines
  represent the now-horizontal continuum level), we reflect the blue
  half of the \halpha\ profile across the peak flux, yielding the
  short-dashed lines. The dotted vertical line is the systemic
  velocity of each object.}\label{f:reflect1}
\end{figure*}

\begin{figure*}
\centering
\centering$
\begin{array}{cc}
\includegraphics[width=3in]{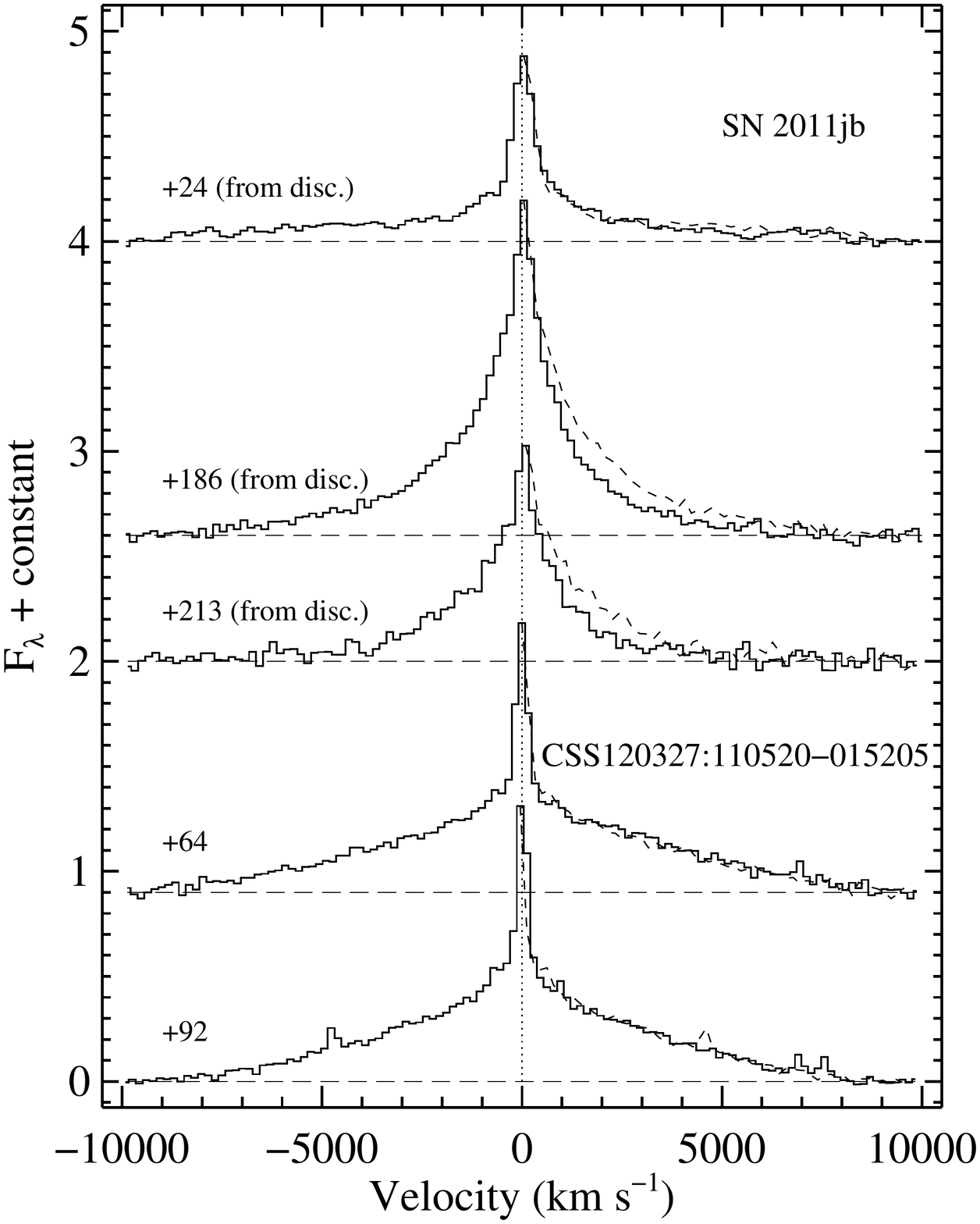} &
\includegraphics[width=3in]{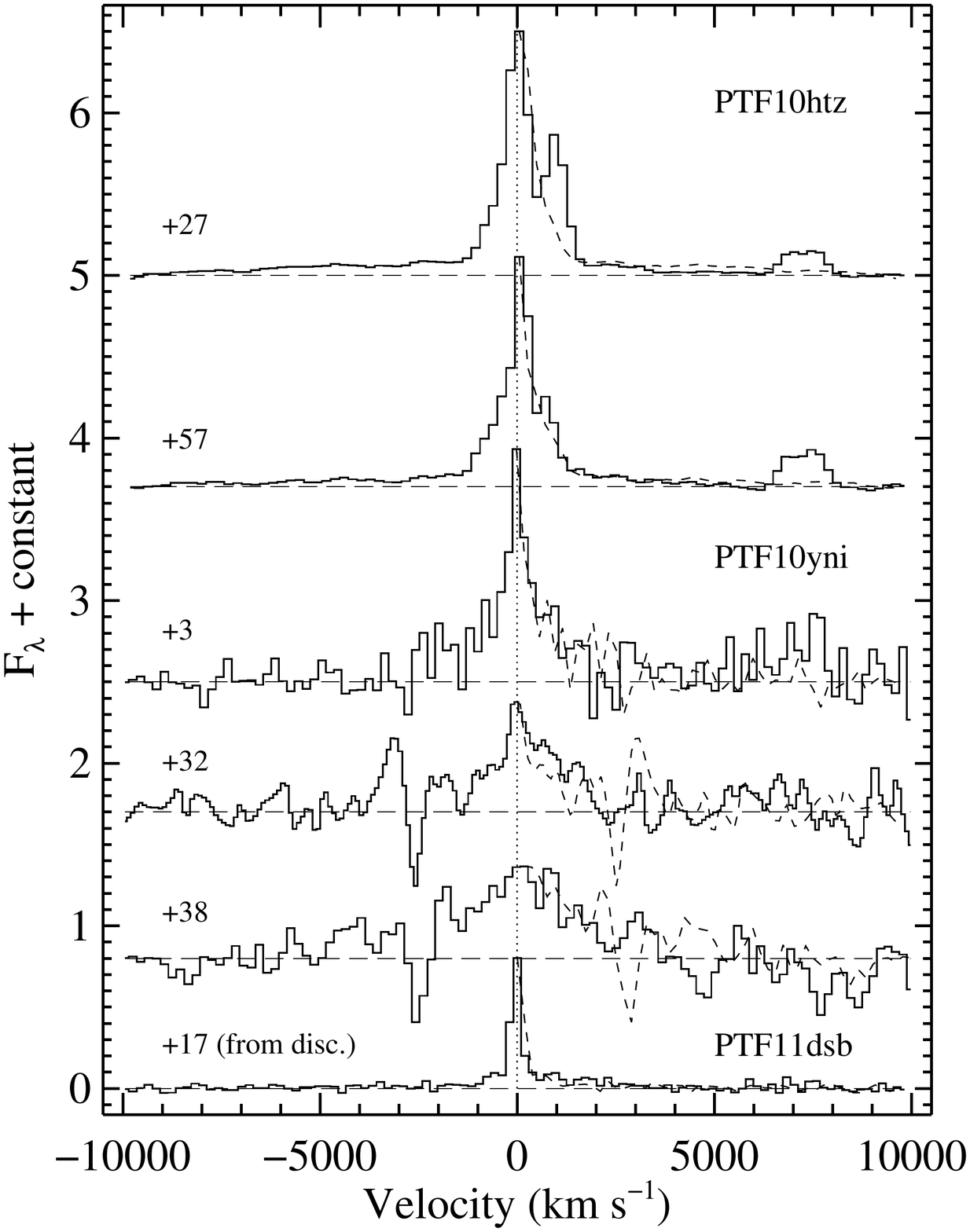} \\
\includegraphics[width=3in]{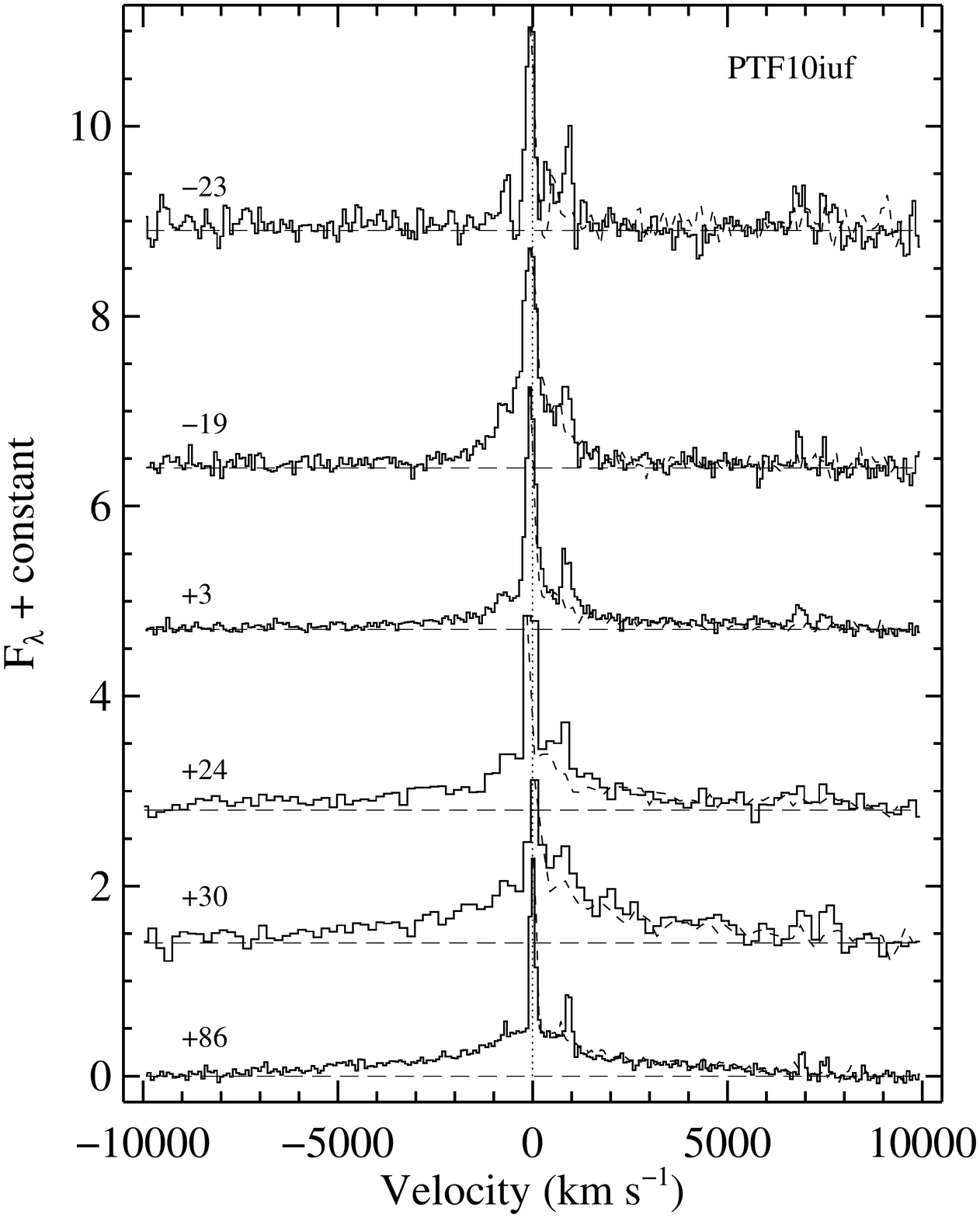} &
\includegraphics[width=3in]{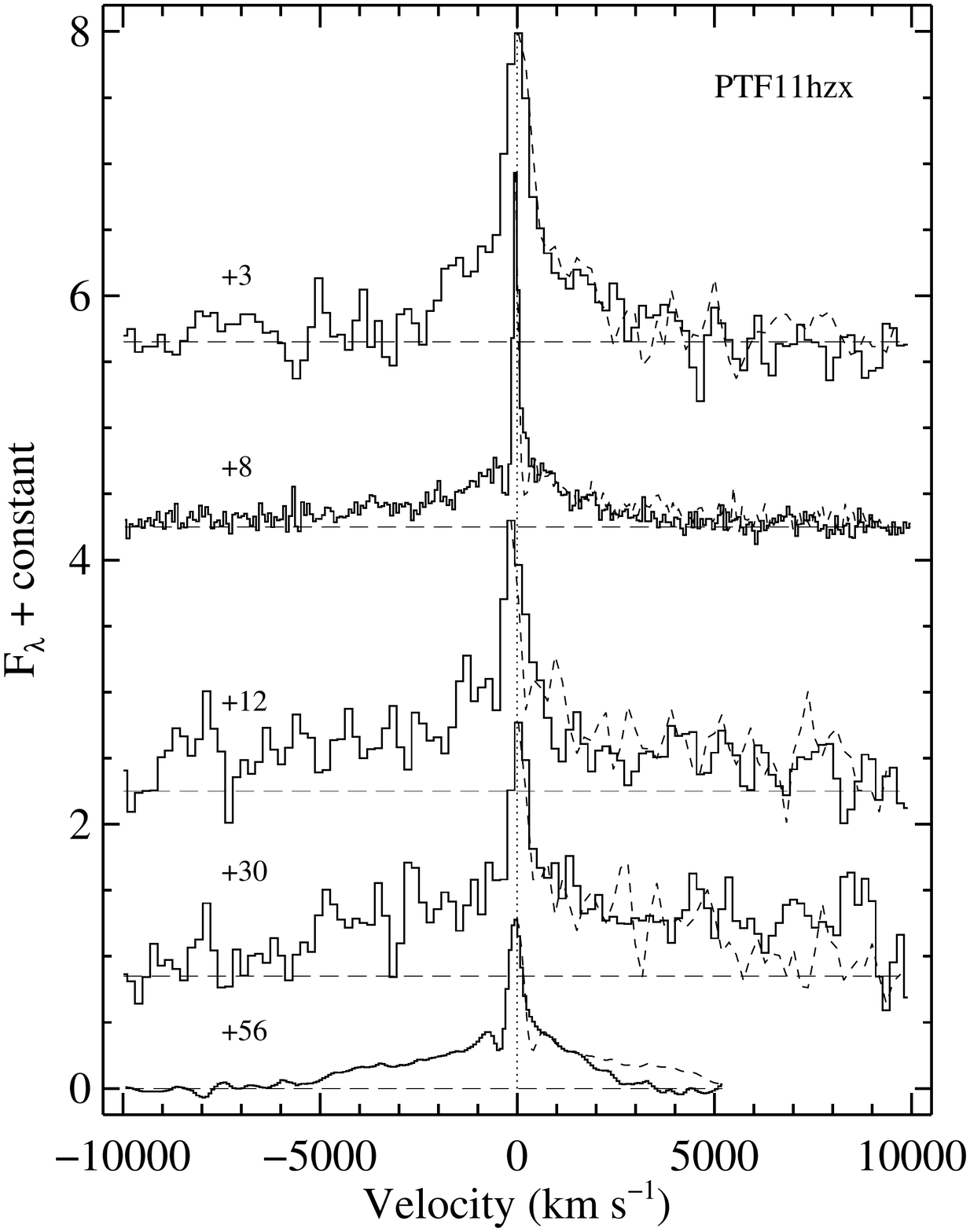} \\
\end{array}$
\caption{The \halpha\ profiles of SN~2011jb, CSS120327:110520--015205,
  PTF10htz, PTF10yni, PTF10iuf, and PTF11hzx. After removing
  a linear 
  continuum (the long-dashed lines 
  represent the now-horizontal continuum level), we reflect the blue
  half of the \halpha\ profile across the peak flux, yielding the
  short-dashed lines. The dotted vertical line is the systemic
  velocity of each object.}\label{f:reflect2}
\end{figure*}

\begin{figure*}
\centering
\centering$
\begin{array}{cc}
\includegraphics[width=3in]{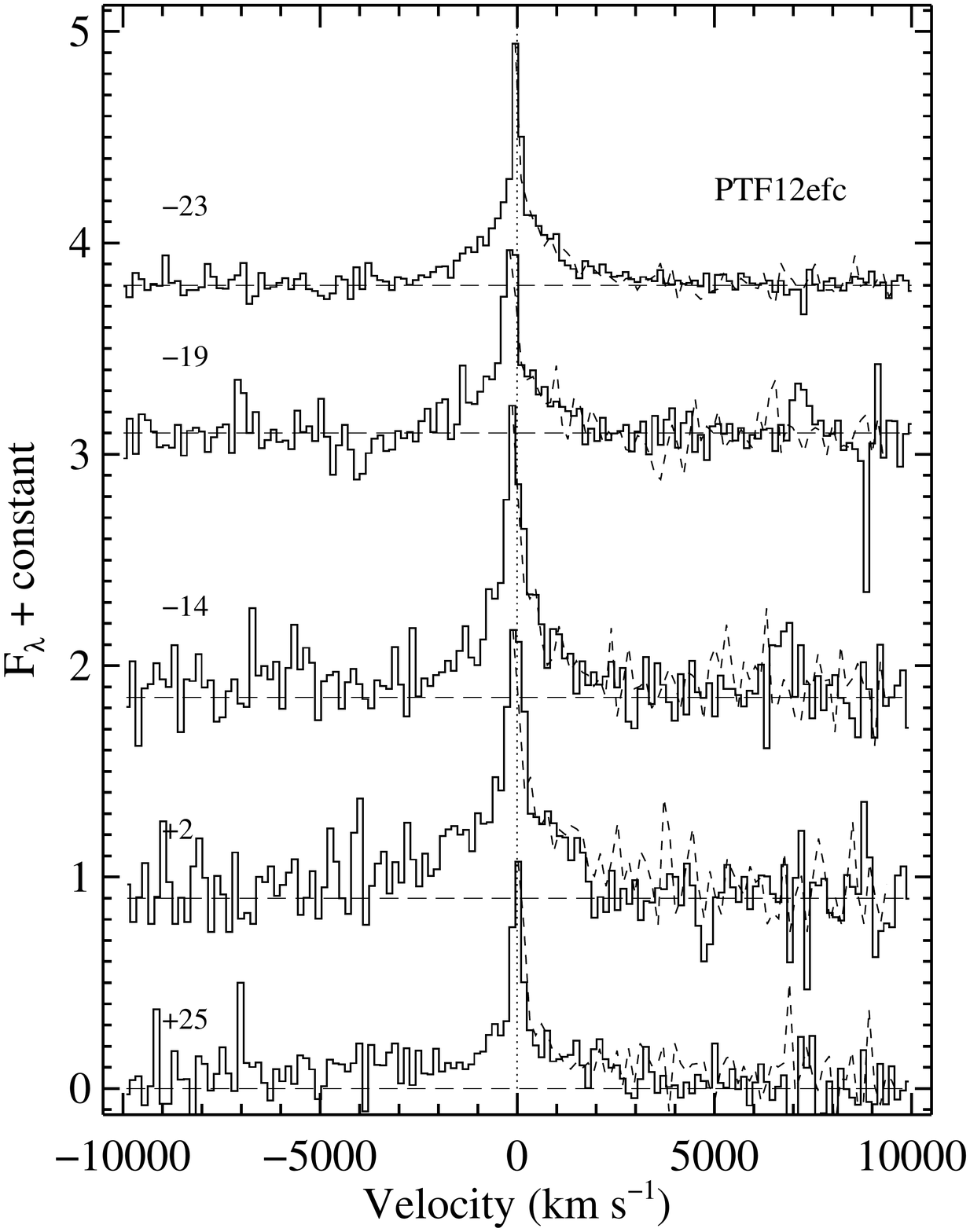} &
\includegraphics[width=3in]{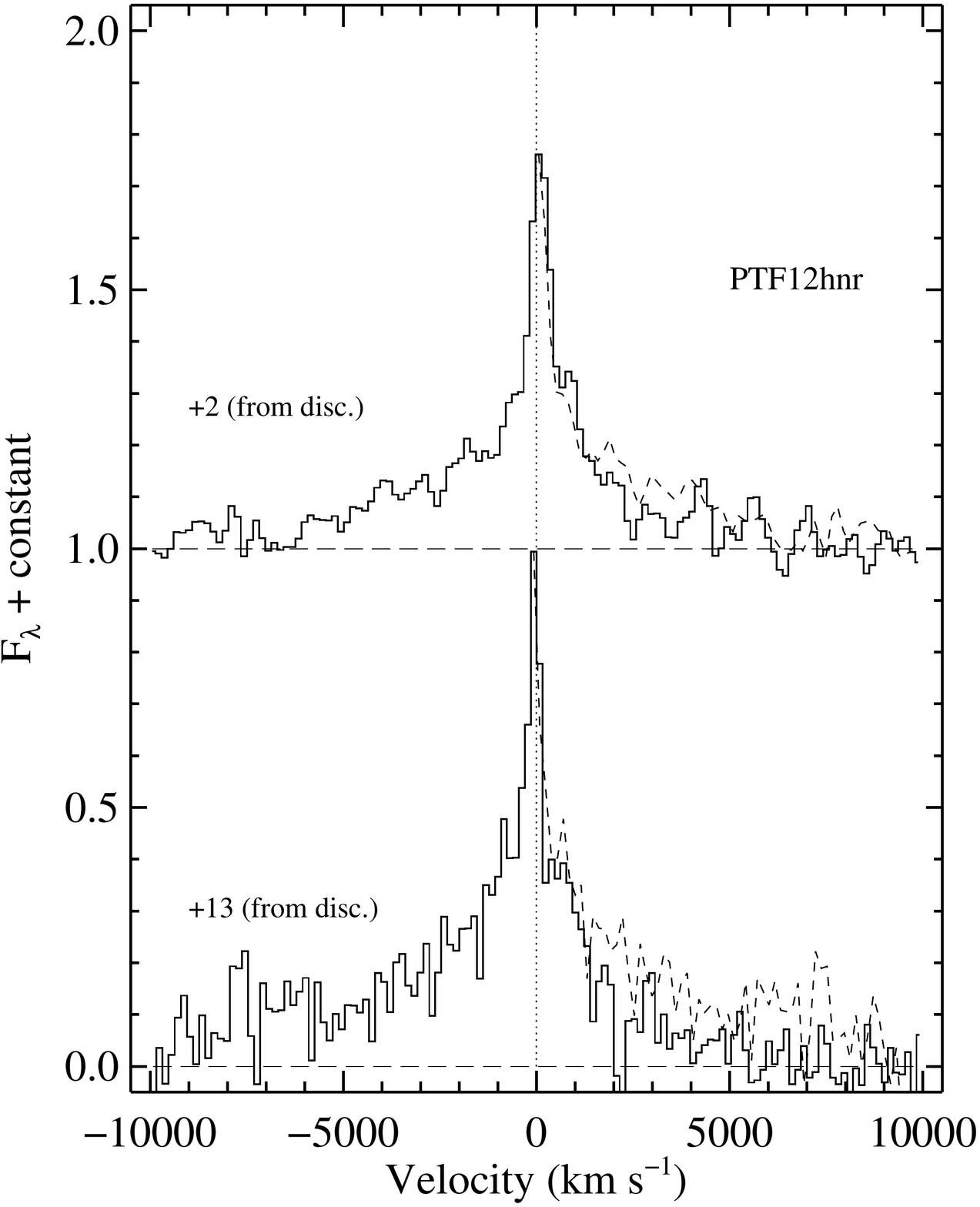} \\
\end{array}$
\caption{The \halpha\ profiles of PTF12efc and PTF12hnr. After
  removing a linear 
  continuum (the long-dashed lines 
  represent the now-horizontal continuum level), we reflect the blue
  half of the \halpha\ profile across the peak flux, yielding the
  short-dashed lines. The dotted vertical line is the systemic
  velocity of each object.}\label{f:reflect3}
\end{figure*}

In more typical SNe~IIn (i.e., ones that almost certainly came from the
core-collapse of a massive star), decreased flux in the red wing of
\halpha\ compared with the blue wing has been interpreted as a sign
of new dust forming in the post-shock material
\citep[e.g.][]{Fox11,Smith12}. In fact, \citet{Fox11} showed this
exact phenomenon for SNe~2008J and
2008cg. Figures~\ref{f:reflect1}--\ref{f:reflect3} display the \halpha\ 
profiles of the 12 SNe~Ia-CSM for which we present spectra in this
paper. Following the method of \citet{Fox11}, we first remove a linear
continuum near \halpha\ from each spectrum (the long-dashed lines in
the figures represent the now-horizontal continuum level). Next we
reflect the blue half of the \halpha\ profile across the peak flux,
yielding the short-dashed lines in the figures. The dotted vertical
lines are the systemic velocity of each object.

We recover the result of \citet{Fox11} that SNe~2008J and 2008cg both
show diminished flux in the red wing of \halpha\ compared with the
blue wing. Surprisingly, we see the exact same behavior in nearly all
of the other non-PTF SNe~Ia-CSM at similar epochs. The red
wings begin to decrease in flux at \about75--100~d past maximum
brightness and seem to continue decreasing with time. We do note, however, 
that there is some evidence that the red flux in SN~2008J begins
to increase again at the latest epochs covered by our spectra. We do
not see this behavior in the PTF SNe~Ia-CSM because we lack spectra at
epochs later than \about75~d past maximum brightness (except for
one spectrum of PTF10iuf taken 86~d past maximum, which possibly shows
evidence of a slight decrease in flux in the red wing of \halpha, as
does the spectrum of PTF11hzx taken 55~d past maximum). In
\citet{Fox11}, $10^{-3}$--$10^{-2}$~\msun\ of dust was inferred from
the mid-IR observations of SNe~2008J and 2008cg, but we caution
that there are quite a few assumptions going into the conversion
from mid-IR photometry to dust mass.
Interestingly, PTF11kx does {\it not} show this evidence for dust
formation through 436~d past maximum brightness. However, there is
some evidence of a decrease in the red wing of \halpha\ in the
spectrum from 680~d past maximum (Silverman \etal submitted).

\begin{figure*}
\centering
\centering$
\begin{array}{c}
\includegraphics[width=5in]{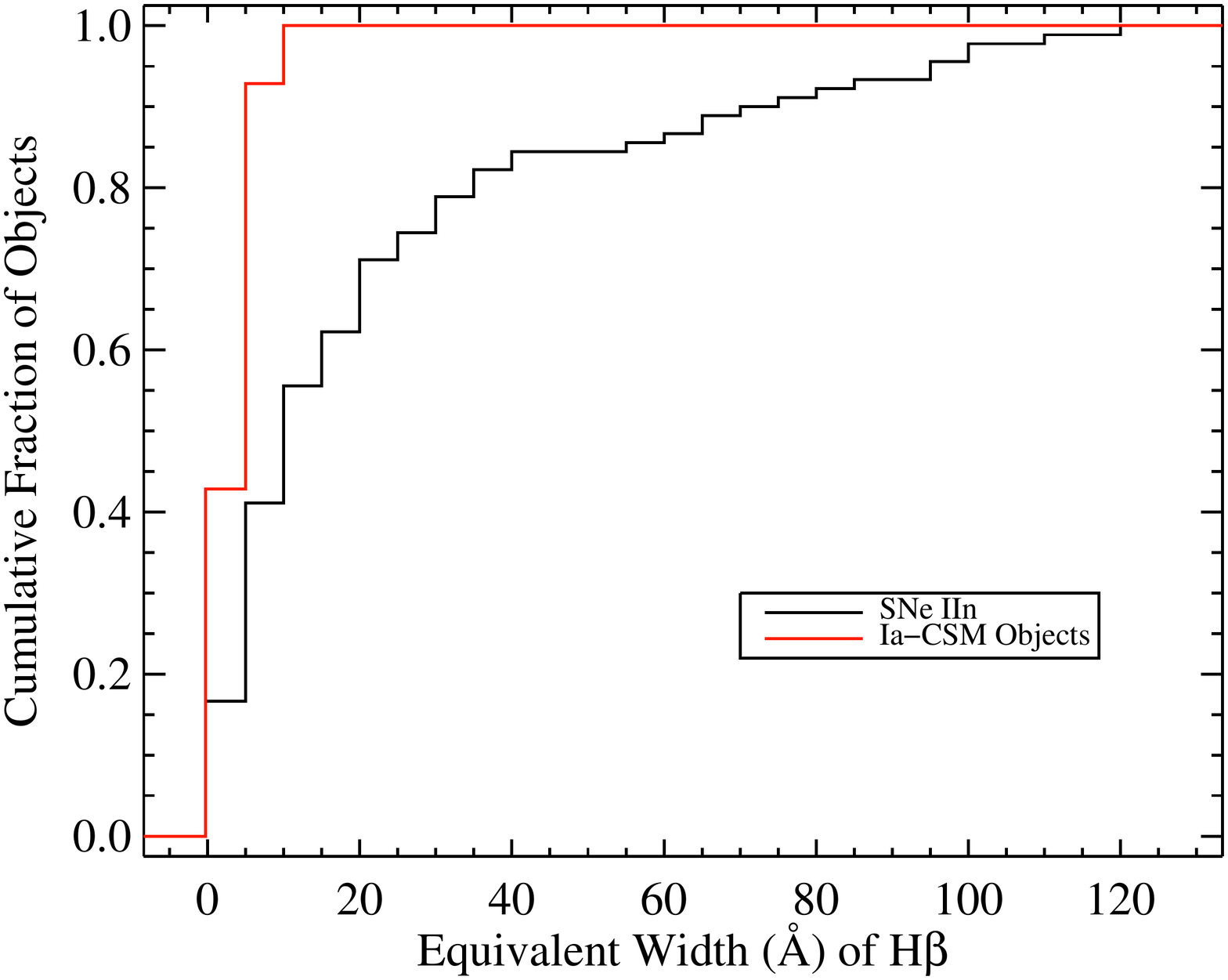} \\
\includegraphics[width=5in]{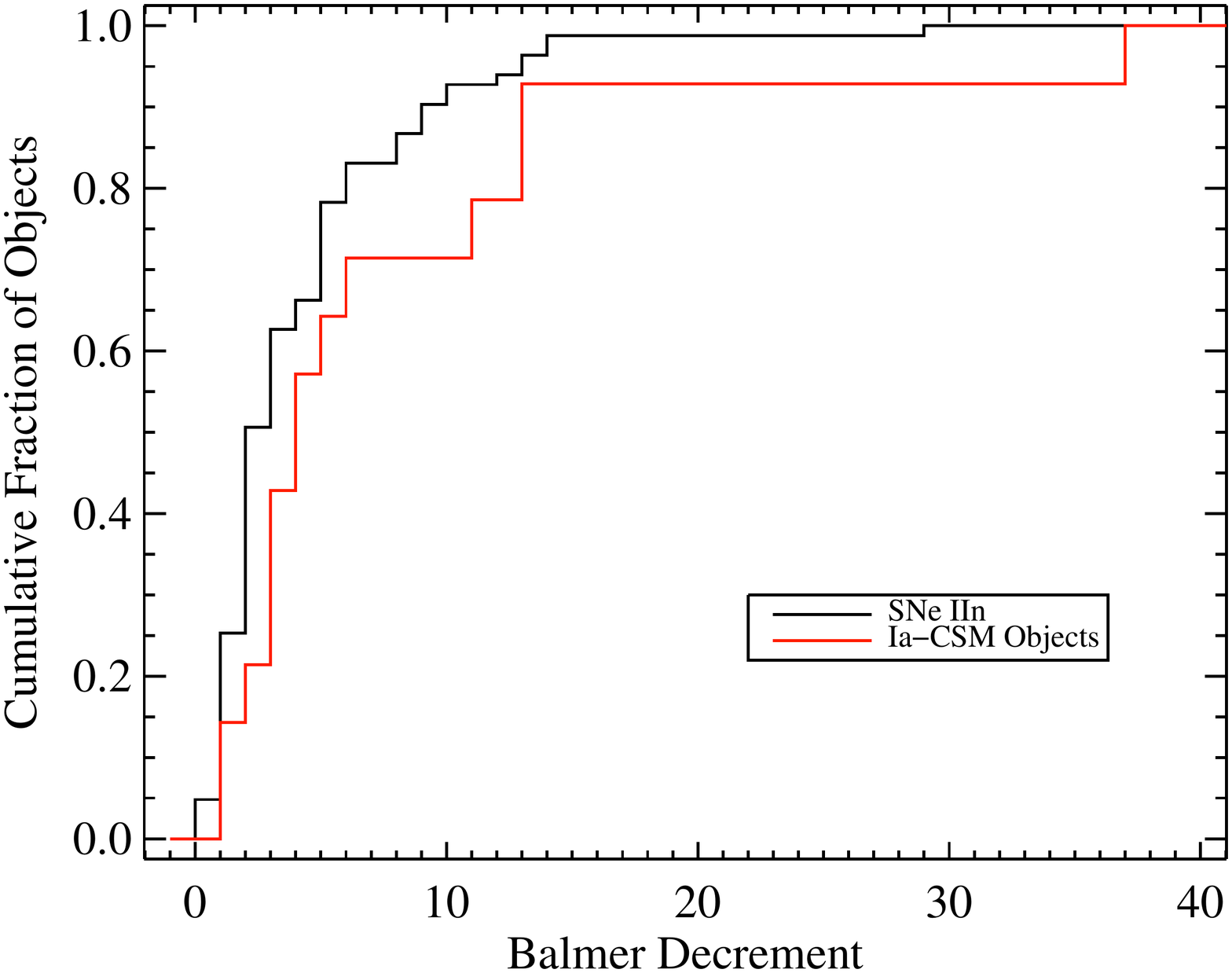} \\
\end{array}$
\caption{Cumulative fraction of \hbeta\ EW ({\it top}) and 
  H$\alpha$/H$\beta$ intensity ratio (Balmer decrement; 
  {\it bottom}) for all SNe~IIn from the Berkeley SN Group's
  database and PTF (black), and for SNe~Ia-CSM with spectra
  presented herein (red). The median values were used when there were
  multiple  
  measurements for a given object.}\label{f:hbeta} 
\end{figure*}

In contrast to \halpha, the EW of \hbeta\ in SNe~Ia-CSM is 
significantly smaller than that of normal SNe~IIn. The median
\hbeta\ EW for SNe~Ia-CSM is \about6~\AA, yet it is
\about13~\AA\ for SNe~IIn. Furthermore, the EW values of the two
SN types seem to be drawn from different parent populations; a
Kolmogorov-Smirnov (KS) test yields $p \approx 0.0003$. The top panel
of Figure~\ref{f:hbeta} shows the cumulative fraction of \hbeta\ EWs
for the SNe~IIn from the Berkeley SN Group's database as well as the
SNe~IIn from PTF. The SNe~Ia-CSM with spectra presented in this
work are represented by the red line while the SNe~IIn are represented
as the black line. The median EW was used when there were multiple EW
measurements for a given object.

Similarly, we can calculate the H$\alpha$/H$\beta$ intensity ratio
(hereafter, the Balmer decrement)
of the SNe~IIn and SNe~Ia-CSM, which is shown in the bottom panel of
Figure~\ref{f:hbeta}. The Balmer decrement is only somewhat smaller in 
SNe~IIn compared with SNe~Ia-CSM (\about3 versus \about5, respectively), 
but according to a KS test this is significant
($p \approx 0.024$). The Balmer decrement in SNe~Ia-CSM also
appears to increase with time before eventually decreasing, peaking at
ages anywhere from about a few months to a year and a half past
maximum brightness. PTF11kx had H$\alpha$/H$\beta > 7$ in all of
its late-time spectra, and it too showed an increase and subsequent 
decrease, achieving a peak value at \about1~yr past
maximum brightness (Silverman \etal submitted). Furthermore, large Balmer 
decrements were also observed in SN~2005gj \citep{Aldering06}.

A possible explanation for these large Balmer decrements
is that the emission lines are produced primarily through collisional
excitation rather than recombination. Recombination should lead to a 
Balmer decrement of \about3, while moderately high density gas can 
lead to large Balmer decrements when the optical depth of \halpha\ is 
large \citep{Drake80}, possibly caused by Balmer self-absorption and
collisional excitation \citep{Xu92}. 
In SNe~Ia-CSM, the SN ejecta may be
interacting with thin, relatively dense, slowly moving shells of CSM
that have cavities on either side of the shell, as one would expect
from recurrent nova eruptions \citep[and as was suggested for
PTF11kx;][]{Dilday12}. When the rapidly moving SN ejecta catch up to
and overtake the more slowly moving thin, dense shells, the hydrogen 
may get collisionally excited and naturally lead to the large
H$\alpha$/H$\beta$ ratios observed in the spectra of SNe~Ia-CSM. 
Furthermore, models of SNe~Ia-CSM interacting with a wind having a
constant mass-loss rate appear to be inconsistent with late-time
photometric observations \citep{Chugai04b}. Thus, this relatively
large Balmer decrement, which is ubiquitous in SNe~Ia-CSM, is
likely caused by interaction with multiple thin, dense shells of CSM.

\begin{figure*}
\centering
\centering
\includegraphics[width=5in]{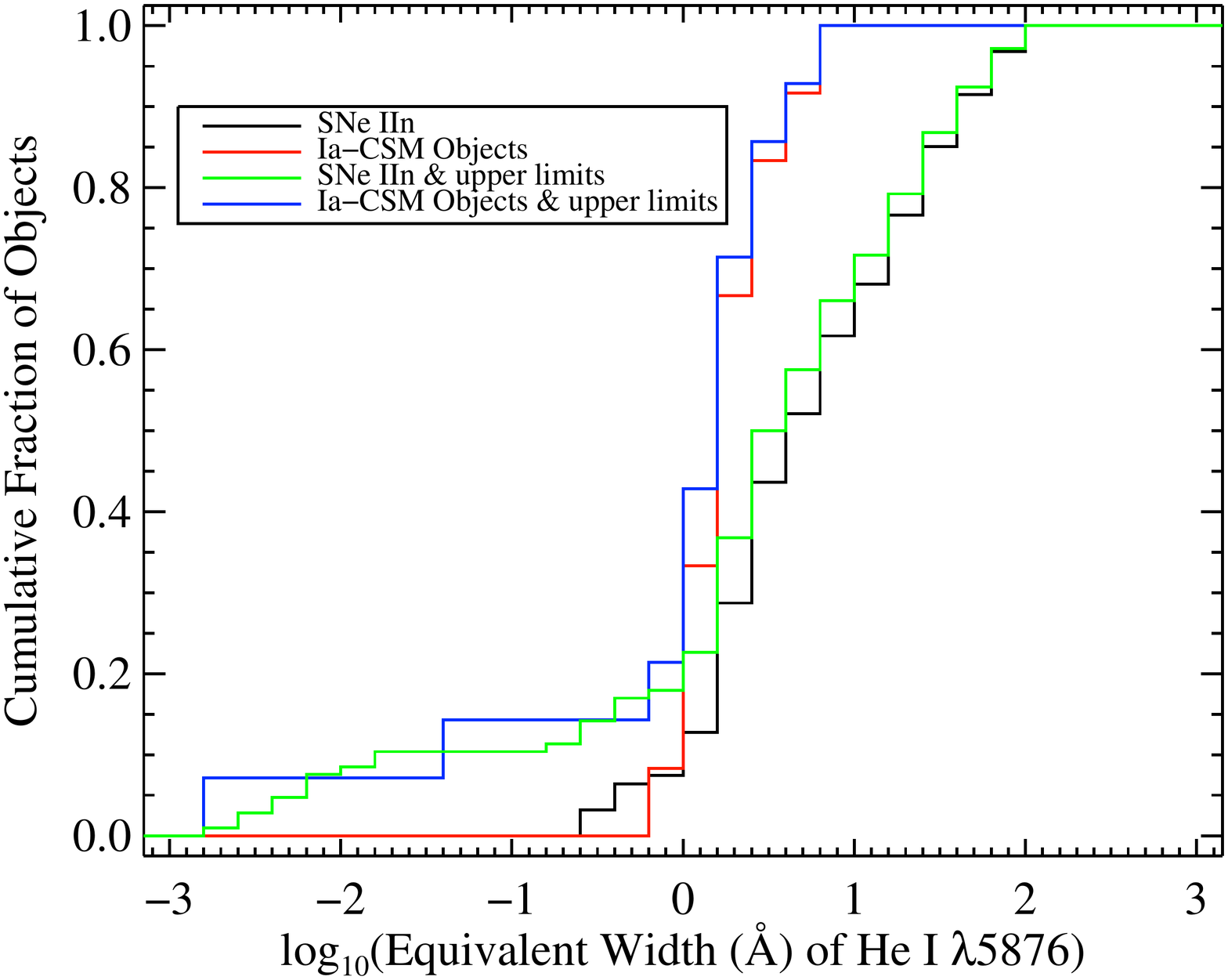}
\caption{Cumulative fraction of \ion{He}{1} $\lambda$5876 EW for all SNe~IIn
  from the Berkeley SN Group's database and PTF (black), and for SNe~Ia-CSM
   with spectra presented herein (red). Also shown are the
  \ion{He}{1} $\lambda$5876 EWs for SNe~Ia-CSM including upper
  limits (blue) and for SNe~IIn as well (green). The median EW value
  is used when there are multiple 
  measurements for a given object.}\label{f:helium}
\end{figure*}

The EW of \ion{He}{1} $\lambda$5876 was relatively small in PTF11kx 
(\about9~\AA; Dilday et al. 2012; Silverman \etal submitted), smaller than 
that of most normal SNe IIn. Weak \ion{He}{1} emission extends beyond 
just PTF11kx to the rest of the SNe~Ia-CSM as well. Relatively narrow
\ion{He}{1} emission was detected in optical and near-IR spectra of
SN~2008J taken a few days before maximum brightness, but the features
were weak \citep{Taddia12}. Figure~\ref{f:helium} shows the
cumulative fraction of EW of \ion{He}{1} $\lambda$5876 for the
Berkeley Group's SNe~IIn and the PTF SNe~IIn in black, and for
SNe~Ia-CSM which have spectra shown in this work in red. The
blue line is the SNe~Ia-CSM, but including upper limits to the
\ion{He}{1} $\lambda$5876 EW, and the green line includes the upper
limits for the {\it non}-SNe~Ia-CSM. As above, the median EW is used
when multiple EW measurements exist for a given object. 

A KS test shows that the difference between SNe~Ia-CSM
and SNe~IIn is significant ($p \approx 0.009$), and the
difference in median EW is striking (\about2~\AA\ versus \about6~\AA,
respectively). Including the upper limits calculated for the EW of
\ion{He}{1} $\lambda$5876, the difference is decreased (median EW
values of \about2 and \about4~\AA\ for SNe~Ia-CSM and SNe~IIn,
respectively), and it is less significant ($p \approx
0.017$). Thus, it seems that little to no \ion{He}{1} emission is a
common attribute of members of the SN~Ia-CSM class. Perhaps the
stronger \ion{He}{1} emission in SNe~IIn is due to an actual abundance
enhancement in the wind from a massive star, as compared to the CSM
shells coming from systems containing a WD
\citep[e.g.,][]{Chevalier94}.

As final note on optical spectroscopy of SNe~Ia-CSM, we find that
in most objects and at nearly all epochs the second strongest feature
(after \halpha) is broad (\about10,000~\kms) emission from the
\ion{Ca}{2} near-IR triplet. Every SN~Ia-CSM discussed herein
(with spectra that encompass this spectral feature near
8500--8600~\AA) shows strong emission from the \ion{Ca}{2} near-IR
triplet (see Figures~\ref{f:sn2005gj}--\ref{f:12hnr}). The feature
also appears prominently in late-time spectra of PTF11kx, though it
almost completely disappears by 680~d past maximum brightness
(Silverman \etal submitted). According to at least one model, this
behavior may be the result of a cool, dense shell with separate Fe-poor
and Fe-rich zones becoming fully mixed \citep{Chugai04}.

\subsection{Other Observations of SNe~Ia-CSM}\label{ss:other_observations}

In addition to optical photometry and low-resolution optical
spectroscopy, there has been a smattering of other types of
observations of SNe~Ia-CSM at a variety of
wavelengths. High-resolution optical spectra exist of relatively few
SNe, and this is true of SNe~Ia-CSM as well. However, for all
SNe~Ia-CSM with published high-resolution spectroscopy, narrow
(50--100~\kms) P-Cygni profiles have been observed in \halpha, as well
as a handful of other spectral features 
\citep[e.g.,][]{Kotak04,Aldering06,Dilday12}. Unfortunately, the
spectra of SNe~Ia-CSM presented herein are of such low resolution
that we do not expect to observe these subtle features.

Weak, narrow \ion{Na}{1}~D absorption from the host galaxy of
SN~2005gj was found in the high-resolution optical spectra presented
by \citet{Aldering06}. On the other hand, \citet{Dilday12} show strong
\ion{Na}{1}~D absorption from the host galaxy of PTF11kx. Furthermore,
a statistical study of many high-resolution optical spectra of SNe was
undertaken by \citet{Sternberg11}, and the two SNe~Ia-CSM in their
sample (SNe~2008J and 2008cg, referred to
as CCSNe) are found to have {\it saturated} \ion{Na}{1}~D lines,
though the lines were redshifted in SN~2008J and blueshifted in
SN~2008cg.

We inspect the low-resolution spectra presented in this work and
find that all SNe~Ia-CSM have likely \ion{Na}{1}~D absorption
from the host galaxy. Regarding the two SNe~Ia-CSM for which we do
not present spectra herein, SN 1997cy
probably does not show \ion{Na}{1}~D absorption from the host
\citep{Turatto00,Germany00}, while SN 2002ic likely does
\citep{Benetti06}. Thus, narrow \ion{Na}{1}~D absorption from the host
galaxy is present in nearly all SNe~Ia-CSM.

High-resolution spectroscopy of SNe may be rare, but SN~2002ic is the
only Ia-CSM object that has published spectropolarimetric
observations. \citet{Wang04} find that this object is mostly evenly
polarized across the optical region at the \about0.8\% level, except 
near \halpha\ where it is highly depolarized. They state that this
is significantly different than more normal SNe~Ia, and that in
the case of SN~2002ic the polarization is likely more closely tied 
to the CSM than to the explosion itself.

At longer wavelengths than optical, we have already pointed out (in
\S\ref{ss:08j_08cg}) that SNe~2008J and 2008cg are detected in the
mid-IR by \citet{Fox11}. In fact, at \about620~d past maximum
brightness, SN~2008J is the second strongest IR detection in their
study, with flux densities of 2.52~mJy and 2.71~mJy at 3.6~$\mu$m and 
4.5$\mu$m, respectively. While not as luminous, SN~2008cg at
\about530~d past maximum is also clearly detected with flux densities
of 0.30~mJy and 0.35~mJy at 3.6~$\mu$m and 4.5$\mu$m,
respectively. The two best-studied SNe~Ia-CSM are also found to
have IR excesses at late times. SN~2002ic is easily detected in
$K$-band imaging \about256~d past maximum brightness as well as in
$H$-band and $K$-band imaging \about352~d past maximum
\citep{Kotak04}. From \about40--140~d past maximum, SN~2005gj is
found to be 2--3~mag brighter and to decline more slowly than normal SNe~Ia
and SNe~IIn in \protect\hbox{$JHK_s$} imaging \citep{Prieto07}. These
IR excesses are consistent with the new dust formation scenario
described above in order to explain the decreasing flux in the red
wing of \halpha\ at late times. In
addition, there exists only one published near-IR spectrum (extending to
2.2~$\mu$m) of a Ia-CSM object (SN~2008J), and it shows emission lines 
of the H Paschen and Brackett series, along with likely \ion{Fe}{2}
emission and weak \ion{He}{1} emission \citep{Taddia12}.

Proceeding to even lower energy radiation, SN~1997cy is not detected
at 13 or 20~cm \about400~d after discovery, though the limits
(\about$2 \times 10^{21}$~W~Hz$^{-1}$) are not very restrictive
\citep{Germany00}. Similarly, neither SN~2005gj \citep[at \about39~d
past maximum brightness;][]{05gj_radio} nor SN~2008cg \citep[at
\about61~d past maximum;][]{08cg_radio} are detected at 8.46~GHz with
the Very Large Array. PTF12hnr, observed at 6.1~GHz using the EVLA
\about20~d after discovery, also resulted in a null detection. These
observations can be converted into radio luminosity upper limits of
\about10$^{26}$--10$^{28}$~\ergps~Hz$^{-1}$. This range encompasses
both upper limits and actual detections of more typical SNe~IIn
\citep[e.g.,][]{Germany00,Pooley02,Fox11}.

At higher energies, both SN~2005gj \citep{05gj_xray} and SN~2008cg
\citep{08cg_xray} are easily detected \about39 and \about62~d past
maximum, respectively, by {\it Swift}/UVOT in the ultraviolet, including 
the bluest filter which covers 112--264~nm. PTF11hzx and PTF12efc, also
observed by {\it Swift}/UVOT (the former \about9~d after maximum
brightness and the latter \about29 and 16~d before maximum),
are both well-detected in all observations.

Moving into the X-rays, SN~2005gj was not detected by {\it Swift}/XRT
\about39~d past maximum brightness \citep{05gj_xray} or by {\it
  Chandra}/ACIS \about55~d past maximum \citep{Prieto07}. Similarly,
no detection of SN~2008cg was made by {\it Swift}/XRT \about62~d past
maximum brightness \citep{08cg_xray}. Unsurprisingly, PTF11hzx was
also not detected by {\it Swift}/XRT \about9~d after maximum
brightness. The upper limits on the X-ray luminosity implied by these
null detections are \about10$^{39}$--10$^{43}$~\ergps; observed X-ray
fluxes of SNe~IIn fall in this range
\citep[e.g.,][]{Pooley02,Zampieri05,Immler07}. Using the upper limits
determined from the X-ray nondetections of SNe~Ia-CSM and
equations in \citet{Ofek13}, we calculate mass-loss rate upper limits
of a few times $10^{-1}$~\msun~yr$^{-1}$ (which matches the mass-loss
rates calculated above using the rise times of SNe~Ia-CSM).

\subsection{Host Galaxies of SNe~Ia-CSM}\label{ss:host}

\citet{Prieto07} found that the host galaxies of the first four 
discovered SNe~Ia-CSM (SNe~1997cy, 1999E, 2002ic, and 2005gj) are all
late-type galaxies (dwarf irregulars and late-type spirals) with
star formation likely occurring within the last few hundred Myr. They
also show that, with the exception of the host of SN~1999E, the hosts
have low luminosities ($-19.1 < M_r < -17.6$~mag, similar to the
Magellanic clouds) which implies subsolar metallicities. The host of
SN~1999E, however, is bright in the IR, shows a nuclear starburst, but
is consistent with solar metallicity and Milky Way (MW) luminosity 
\citep{Allen91,Prieto07}.

Using the NASA/IPAC Extragalactic Database (NED) and the Sloan Digital
Sky Survey Data Release 8 \citep[SDSS DR8;][]{SDSSDR8}, we find that
the hosts of the four other non-PTF SNe~Ia-CSM discussed in this
work are also late-type galaxies. Three of the four have relatively
low luminosities ($-19.3 < M_r < -18.1$~mag), while the host of
SN~2008J appears to be consistent with MW luminosity (much like the
host of SN~1999E).

Turning now to the PTF SNe~Ia-CSM, PTF11kx has a
late-type spiral host with luminosity comparable to that of the MW and
slightly higher-than-solar metallicity
\citep{Dilday12,Tremonti04}. Four of the seven newly discovered 
SNe~Ia-CSM in PTF (PTF10htz, PTF10iuf, PTF11dsb, and PTF12hnr) are also
found in what are likely late-type spiral hosts, all of which have
luminosities similar to that of the MW ($-20.6 < M_r < -19.2$~mag). The
remaining three PTF SNe~Ia-CSM (PTF10yni, PTF11hzx, and PTF12efc) do not
have detectable hosts in SDSS DR8 \citep{SDSSDR8} or in our deep
stacks of PTF search images. This implies that they are low-luminosity
galaxies with $M_r \ga -18$~mag.

Thus, of the 16 SNe~Ia-CSM discussed herein, all appear to
have exploded in late-type 
galaxies. Nine of them are found in low-luminosity (and presumably,
low-metallicity) hosts, including three hosts that are not detected by
SDSS DR8 or PTF, while seven of them (including PTF11kx) are found in
hosts roughly similar to the MW. Interestingly, when the hosts of PTF
SNe~Ia-CSM are compared to the hosts of SNe~IIn from PTF, no
significant differences are found. It has been shown in earlier work
that SNe~II in general have statistically the same hosts as
SNe~IIn \citep{Kelly11}, and that SNe~IIn and SNe~IIP both trace recent
(but not ongoing) star formation \citep{Anderson12}. Furthermore,
contrary to some expectations,\citet{Anderson12} find that SNe~IIn do not
come from the youngest (and thus most massive) stars; instead, their 
progenitors are slightly older than those of
SNe~IIP. They also find that the association with host-galaxy \halpha\ 
emission for SNe~IIn is between that of other SNe~II and SNe~Ia, perhaps
implying that some of their SNe~IIn are actually SNe~Ia-CSM which
have lower-mass progenitors than any CCSN subtype.\footnote{Note that
  none of the SNe~Ia-CSM discussed herein were part of the
  \citet{Anderson12} sample.}


\section{Conclusions}\label{s:conclusions}

Running SNID on the Berkeley SN Group's database of SN~IIn spectra, we
have reidentified four SNe~Ia-CSM that were previously known
(SNe~2008J, 2008cg, 2011jb, and CSS120327:110520--015205) but poorly
studied. These 
are in addition to the four well-studied Ia-CSM objects (SNe~1997cy,
1999E, 2002ic, and 2005gj). Furthermore, SNID was run on all 63
SNe~IIn discovered by PTF through August~2012, and seven new SNe~Ia-CSM
were uncovered. Armed with a sample of 15 SNe~Ia-CSM, in
addition to PTF11kx (\citealt{Dilday12}; Silverman \etal
submitted), we investigate the unifying characteristics of this class of
SN. Observable signatures of SNe~Ia-CSM are as follows.

\begin{itemize}

\item Peak absolute magnitudes of $-21.3 \leq M_R \leq -19$~mag are
  observed (somewhat more luminous than normal SNe~Ia 
  and the bulk of the SNe~IIn population, but less luminous than
  superluminous SNe) and relatively long rise times of \about20--40~d 
  (as opposed to \about18~d for more normal SNe~Ia). 

\item SNID cross-correlations of optical spectra show that SNe~Ia-CSM
  are spectroscopically homogeneous. The spectra consist of a
  diluted SN~Ia spectrum, along with a 
  relatively blue ``quasi-continuum'' from many blended lines of IGEs,
  strong and broad (\about10,000~\kms) emission from the \ion{Ca}{2} 
  near-IR triplet, and are dominated by \halpha\ emission with widths of
  \about2000~\kms.

\item The \halpha\ profile shows strong fluctuations until
  \about100--150~d past maximum brightness, at which time the strength
  tends to increase with time. There are also extremely narrow
  (50--100~\kms) P-Cygni profiles present in \halpha, and after 
  \about75--100~d past maximum a decrease in flux in the red wing is
  seen (often attributed to newly formed dust). 



\item Weak \ion{He}{1} and \hbeta\ emission are seen, as compared to
  typical SNe~IIn, and the H$\alpha$/H$\beta$ intensity ratio is large.
  This is likely caused by collisional excitation of hydrogen when the
  SN~Ia-CSM ejecta overtake slower-moving, thin, moderately dense CSM
  shells.




\item Within the first few months after explosion, ultraviolet emission is
  seen, but no radio or X-ray emission is detected (although the range
  of upper limits derived is consistent with both upper limits and
  actual detections of SNe~IIn at radio and X-ray wavelengths). Mid-IR
  emission is visible at \about0.5--2~yr past maximum brightness, and it 
  is stronger than in typical SNe~IIn or SNe Ia (which is further
  evidence of newly formed dust). 



\item Using both rise times and X-ray upper
  limits, SNe~Ia-CSM seem to have mass-loss rates of a few times
  $10^{-1}$~\msun~yr$^{-1}$, though we caution that there are many
  assumptions that go into these calculations.

\item The hosts of SNe~Ia-CSM all appear to be
  late-type spirals with either MW-like luminosities (and solar
  metallicities) or dwarf irregulars having luminosities similar
  to those of the Magellanic Clouds (and thus
  presumably subsolar metallicities).

\end{itemize}

While the above list describes attributes of the Ia-CSM objects,
several of the items can also describe more typical SNe IIn. This
brings up the possibility, noted 
previously \citep[e.g.,][]{Anderson12}, that some subset of the SN~IIn
population is contaminated by SNe~Ia-CSM. Perhaps there are simply
some {\it bona fide} SNe~IIn that share with SNe~Ia-CSM many, but not
all, of the observational characteristics listed above. It is
also plausible that some SNe~IIn that can be described by many of the
items listed above are, in fact, SNe~Ia-CSM where the CSM
interaction is too strong or begins too early for any hint of an
underlying SN~Ia to be identified. On the other hand, at least {\it
  some} SNe~IIn are actual core-collapse events since a massive
progenitor star is detected \citep[e.g., SN~2005gl;][]{Gal-Yam09}.

The existence of the SN~Ia-CSM class of objects seems to argue that at 
least some SNe~Ia arise from the SD channel, since a H-rich
CSM is likely a result of that progenitor scenario \citep[but
see][]{Shen13}. Detailed modeling of both the core collapse of massive
stars and the thermonuclear explosion of WDs (with an assortment of
binary companions) within various arrangements of CSM is beyond the
scope of this paper, but we hope that future theoretical work
on this subject will be informed by the observational results
presented herein.

\begin{acknowledgments}
We would like to thank K. Alatalo, T. Barlow, E. Bellm, 
B. Cobb, A. Cucchiara, M. Ganeshalingam, Y. Green, M. Hidas, 
L. Kewley, N. Konidaris, S. Lazarevic, N. Lee, D. Levitan, 
M. McCourt, K. Mooley, R. Mostardi, D. Perley,
A. G. Riess, B. Sesar, R. Street, T. Treu, V. Viscomi, and X. Wang 
for their assistance with some of the observations and data reduction;
B. Dilday, O. Fox, and L. Wang for helpful discussions; and D. Balam,
M. Stritzinger, J. Vinko, and J. C. Wheeler   
for providing unpublished spectra of possible SNe~Ia-CSM. 
We are grateful to the staffs at the Isaac Newton Group of Telescopes
and the Lick, Keck, Palomar, and Kitt Peak National Observatories for
their support. Some of the data presented herein were obtained at the 
W. M. Keck Observatory, which is operated as a scientific partnership
among the California Institute of Technology, the University of
California, and the National Aeronautics and Space Administration
(NASA); the observatory was made possible by the generous financial
support of the W. M. Keck Foundation. The authors wish to recognize
and acknowledge the very significant cultural role and reverence that
the summit of Mauna Kea has always had within the indigenous Hawaiian
community; we are most fortunate to have the opportunity to conduct
observations from this mountain. This research has made use of the
NASA/IPAC Extragalactic Database (NED) which is operated by the Jet
Propulsion Laboratory, California Institute of Technology, under
contract with NASA. Funding for SDSS-III has been provided by the
Alfred P. Sloan Foundation, the Participating Institutions, the
National Science Foundation, and the U.S. Department of Energy Office
of Science. The SDSS-III web site is http://www.sdss3.org/. 
Supernova research by A.V.F.'s group
at U.C. Berkeley is supported by Gary and Cynthia Bengier, the Richard
and Rhoda Goldman Fund, the Christopher R. Redlich Fund, the TABASGO
Foundation, and NSF grants AST-0908886 and AST-1211916.
Work
by A.G.-Y. and his group is supported by grants from the ISF, BSF, GIF,
Minerva, an FP7/ERC grant, and the Helen and Martin Kimmel Award for
Innovative Investigation. M.M.K. acknowledges generous support from a
Hubble Fellowship and a Carnegie-Princeton Fellowship.
\end{acknowledgments}


\end{document}